\documentclass[aps,pre,superscriptaddress,nofootinbib,preprintnumbers,showkeys]{revtex4-2}

\usepackage{natbib}
\usepackage{graphicx}
\usepackage{graphics}
\usepackage{amssymb}
\usepackage{amsmath,amssymb}
\usepackage[usenames]{color}
\usepackage{subfigure}
\usepackage{hyperref}
\usepackage{breakurl}
\usepackage{enumitem}
\usepackage{float}
\usepackage[normalem]{ulem} 

\def\Tr{\,{\rm Tr}\,}

\renewcommand{\Re}[1]{\hbox{Re}~#1}
\renewcommand{\Im}[1]{\hbox{Im}~#1}

\newcommand{\be}{\begin{equation}}
\newcommand{\ee}{\end{equation}}
\newcommand{\bea}{\begin{eqnarray}}
\newcommand{\eea}{\end{eqnarray}}
\newcommand{\ben}{\begin{enumerate}}
\newcommand{\een}{\end{enumerate}}
\newcommand{\bit}{\begin{itemize}}
\newcommand{\eit}{\end{itemize}}

\newcommand{\la}[1]{\label{#1}}

\newcommand{\eq}[1]{eq.~(\ref{#1})}

\newcommand{\Eq}[1]{Eq.~(\ref{#1})}

\newcommand{\Sec}[1]{Sec.~\ref{#1}}

\newcommand{\Fig}[1]{Fig.~\ref{#1}}

\def\a{\alpha}

\def\nl{\nonumber \\}
\def\nn{\nonumber}

\newcommand\La{{\cal L}}

\newcommand{\bert}{\raise-0.45mm\hbox{\Large$\Box$}}			

\newcommand{\gd}{\gamma_\downarrow}						
\newcommand{\gu}{\gamma_\uparrow}						

\makeatletter
\newcommand*\bigcdot{\mathpalette\bigcdot@{.5}}
\newcommand*\bigcdot@[2]{\mathbin{\vcenter{\hbox{\scalebox{#2}{$\m@th#1\bullet$}}}}}
\makeatother

\definecolor{BrickRed}{cmyk}{0,0.89,0.94,0.28}					
\definecolor{MidnightBlue}{cmyk}{0.98,0.13,0,0.43}				
\definecolor{DarkGreen}{rgb}{0.100806,0.495968,0.209979}
\definecolor{orange}{rgb}{0.587167,0.354498,0.146197}

\begin{document}

\title{The Josephson junction as a quantum engine}

\author{Robert Alicki}
\email{robert.alicki@ug.edu.pl}
\affiliation{International Centre for Theory of Quantum Technologies (ICTQT), University of Gda\'nsk, 80-308, Gda\'nsk, Poland}
\author{Micha\l~Horodecki}
\email{michal.horodecki@ug.edu.pl}
\affiliation{International Centre for Theory of Quantum Technologies (ICTQT), University of Gda\'nsk, 80-308, Gda\'nsk, Poland}
\email{michal.horodecki@ug.edu.pl}
\author{Alejandro Jenkins}
\email{alejandro.jenkins@ug.edu.pl}
\affiliation{International Centre for Theory of Quantum Technologies (ICTQT), University of Gda\'nsk, 80-308, Gda\'nsk, Poland}
\affiliation{Laboratorio de F\'isica Te\'orica y Computacional, Escuela de F\'isica, Universidad de Costa Rica, 11501-2060, San Jos\'e, Costa Rica}
\author{Marcin \L obejko}
\email{marcin.lobejko@ug.edu.pl}
\affiliation{International Centre for Theory of Quantum Technologies (ICTQT), University of Gda\'nsk, 80-308, Gda\'nsk, Poland}
\author{Gerardo Su\'arez}
\email{gerardo.suarez@phdstud.ug.edu.pl}
\affiliation{International Centre for Theory of Quantum Technologies (ICTQT), University of Gda\'nsk, 80-308, Gda\'nsk, Poland}

\date{First public version: 9 Feb.\ 2023.  This revision: 6 Nov.\ 2023.  To appear in {\it New Journal of Physics}}

\begin{abstract}
We treat the Cooper pairs in the superconducting electrodes of a Josephson junction (JJ) as an open system, coupled via Andreev scattering to external baths of electrons.  The disequilibrium between the baths generates the direct-current bias applied to the JJ.  In the weak-coupling limit we obtain a Markovian master equation that provides a simple dynamical description consistent with the main features of the JJ, including the form of the current-voltage characteristic, its hysteresis, and the appearance under periodic voltage driving of discrete Shapiro steps.  For small dissipation, our model also exhibits a self-oscillation of the JJ's electrical dipole with frequency $\Omega = 2 e V / \hbar$ around mean voltage $V$.  This self-oscillation, associated with ``hidden attractors'' of the nonlinear equations of motion, explains the observed production of monochromatic radiation with frequency $\Omega$ and its harmonics.  We argue that this picture of the JJ as a quantum engine resolves open questions about the Josephson effect as an irreversible process and could open new perspectives in quantum thermodynamics and in the theory of dynamical systems.
\end{abstract}


\maketitle

\tableofcontents

\section{Introduction}
\la{sec:intro}

In 1962, Josephson predicted that two superconducting electrodes separated by a thin insulating barrier should exhibit interesting properties associated with the tunneling through that barrier of Cooper pairs (bound states of two electrons, whose condensation at low temperatures accounts for superconductivity) \cite{Josephson}.  The tunneling current through such a ``Josephson junction'' (JJ) is usually expressed as
\be
I_J = I_c \sin \phi
\la{eq:DC}
\ee
where $I_c$ is a constant (the ``critical current'') corresponding to the maximum value of the supercurrent that the JJ can support and $\phi$ is the difference in the phases of the Ginzburg-Landau wavefunctions for each of the electrodes.  If a direct-current (DC) bias is applied, the phase varies in time as
\be
\dot \phi = \frac{2 e V}{\hbar} .
\la{eq:dphi}
\ee
where $V$ is the voltage jump across the junction.  Combining Eqs.\ \eqref{eq:DC} and \eqref{eq:dphi}, the constant bias $V$ should generate an alternating current (AC) through the junction of the form
\be
I_J(t) = I_c \, \sin \left( \phi_0 + \frac{2 e V t}{\hbar} \right) .
\la{eq:AC}
\ee
Josephson's main predictions were soon confirmed by experiment \cite{Anderson, Shapiro} and the JJ has since been an active subject of research in both pure and applied physics; see \cite{Barone, Likharev, Wolf, Tafuri} and references therein.

The JJ's conversion of a constant bias $V > 0$ into an alternating current $I_J$ (we refer to such a process as \hbox{``DC $\to$ AC''}) implies that the JJ should be understood as an electrical {\it self-oscillator} \cite{SO}.  The first practical \hbox{DC$\to$AC} converter was built by Hertz in 1887 using a spark gap (two conductors separated by a thin gap filled with ambient air), allowing him to produce electrical oscillations with frequencies $\sim 100$ MHz \cite{Hertz}.  Hertz used this spark-gap oscillator to study the properties of electromagnetic waves.  The spark gap then became the basis of radio technology, until it was replaced by the related ``singing arc'' of Duddell \cite{Duddell}, and later by triode self-oscillators (see, e.g., \cite{Groszkowski, Ginoux} and references therein).  These developments motivated a sophisticated mathematical theory of classical self-oscillation by Poincar\'e, Andronov, Hopf, van der Pol, and others.  That approach has been based principally on the theory of ordinary differential equations and of nonlinear phenomena.\footnote{For an interesting discussion of Poincar\'e's seminal contribution to this subject, its connection to the singing arc problem, and the relation to subsequent work by other mathematical scientists, see \cite{Ginoux}.  On the early history of the singing arc and its applications, as well as its connection with the modern concept of ``memristor'', see \cite{arc1, arc2}.}

Meanwhile the subject of DC$\to$AC and of self-oscillation more generally has attracted relatively little attention in theoretical physics.  According to Le Corbeiller,
\begin{quote}  
there is some danger that the beauty of these kinematic developments could lead the student or reader into forgetting the physical relationships in the actual {\it oscillator}, the energy exchanges in particular, as Charles Fabry pointed out to the writer about 1929.  Papers \cite{LeCorbeiller-Fr, LeCorbeiller-Eng} were inspired by this remark. \cite{2Stroke}
\end{quote}
The present work pursues, in a quantum context, the program suggested long ago by Fabry and Le Corbeiller and advanced more recently by some of the present authors of describing self-oscillators as {\it engines} that generate work cyclically and irreversibly, at the expense of an external thermodynamic disequilibrium \cite{SO, engines}.

Motivated by these considerations, in this article we propose a simple model of the JJ as an {\it open quantum system}, consisting of the superconducting Cooper pairs, weakly coupled to two external baths of electrons (each bath corresponds to a terminal of the battery that provides the DC bias).  Our description is based on the Markovian master equation (MME) for the irreversible dynamics of the open system.  Our model modifies the form of the relations between the JJ's tunneling current, voltage, and phase that appear in the literature [Eqs.\ \eqref{eq:DC}, \eqref{eq:dphi}, and \eqref{eq:AC}], while giving the right properties for the JJ's current-voltage characteristic, including the hysteresis seen when the dissipation is small and the ``Shapiro steps'' that appear when the JJ is placed in an oscillating electric field.

The nonlinear dynamical equations resulting from our model also exhibit {\it self-oscillations} about stable fixed points on the JJ's current-voltage characteristic.  For small dissipation and significant bias voltage $V$, we find that a finite but small perturbation can take the system out of the fixed point's basin of attraction, allowing the JJ's electrical dipole to self-oscillate with frequency $\Omega = 2 e V / \hbar$.  This sort of structure has been called a ``hidden attractor'' in the theory of dynamical systems \cite{hidden}.  We will argue that this provides a picture, consistent with the principles of thermodynamics, of how the JJ can act as an {\it engine} that cyclically generates work in the form of monochromatic radiation at the frequency $\omega$ and its harmonics.

Such non-thermal radiation from a JJ has been observed experimentally, both as electromagnetic waves (photons) \cite{Langenberg, Yanson, Pedersen, THz-review, laser} and as sound (phonons) \cite{phonons, decoherence}.  The emission of photons by the JJ is enhanced when the device is coupled to a ``Fiske resonance'' (as in the setups described in \cite{Langenberg, Yanson, THz-review}), or placed inside a resonant cavity (as in \cite{Pedersen, laser}), but the JJ itself must exhibit a self-oscillating engine dynamics in order to account for the monochromatic photon emission.  That the JJ is an autonomous engine, which extracts work from the disequilibrium of the external baths, is also demonstrated experimentally by the fact that the JJ's rotating phase can be used to pump a current along another circuit, as demonstrated experimentally in \cite{JJ-pump}.

Textbooks in electronics distinguish between {\it active} devices that can amplify the power that they receive from the circuit, and {\it passive} devices that cannot.  Horowitz and Hill note that active devices are characterized ``by their ability to make oscillators, by feeding from output signal back into the input,'' i.e., to self-oscillate \cite{HH}.  In this classification, resistors, capacitors, and inductors are passive, whereas triodes, transistors, operational amplifiers, and JJ's are active.

We stress that in this work we seek to understand how the JJ {\it itself} acts as a (chemical) engine.  This is different from other recent work in quantum thermodynamics in which the JJ has been considered as an element for building larger engines.  For instance, in \cite{jj-engine1,jj-engine2} the authors proposed thermal machines based on circuits that include a JJ coupled to cavities.  In this article we will not consider such larger devices, but will instead work out a simple open-system model for the JJ, allowing us to describe the Josephson effect as an irreversible process.

The plan for this article is as follows: In \Sec{sec:model} we start by reviewing the existing theoretical descriptions of the JJ, based on the two-component macroscopic wave-function and the ``resistively and capacitively shunted junction'' (RCSJ).  We will argue that these treatments are not quite consistent.  Instead, one should regard the JJ as a quantum engine, in which the ``working substance'' of the Cooper-pair condensates in the JJ's electrodes extracts work cyclically from the external disequilibrium between two baths of electrons.  Electrons from a bath can combine to form a Cooper pair in the system, and a Cooper pair from the system may decay into two electrons in a bath.  This system-bath interaction corresponds to the well known process of Andreev scattering \cite{Andreev}.  We describe the dynamics of this open system using a Markovian master equation (MME).  From this MEE we obtain coupled ordinary differential equations that describe the evolution in time of the JJ's voltage $V$ and of the coherence $z$ of the JJ considered as a macroscopic two-level quantum system.

In \Sec{sec:characteristic} we study the equilibria of these equations of motion and their linear stability, obtaining a current-voltage characteristic whose features are consistent with the observed properties of a JJ.  In \Sec{sec:SO} we study the dynamics of the model, and find that it exhibits ``hidden attractors'' that can account for the self-oscillation of the electric dipole that causes a DC-biased JJ to emit monochromatic radiation at the Josephson frequency $\Omega = 2 e V / \hbar$.  The details of how this radiation is produced, and how it is enhanced by coupling to a resonant cavity, are considered in \Sec{sec:radiation}.  This allows us to obtain an expression for the thermodynamic efficiency of the JJ as an engine, with the integrated power of the non-thermal radiation considered as its work output.  We compare the results of our model to experimental observations of emission from JJ's and find qualitative agreement.  Finally, we summarize our results and consider some avenues for further investigation in \Sec{sec:discussion}.

\section{Need for open-system approach} 
\la{sec:need_open}

In this section we will review how the JJ is commonly described in the literature in terms of the Cooper-pair condensate's ``macroscopic wave function'' (MWF) and of the ``resistively and capacitively shunted junction'' (RCSJ) model.  We will argue that these approaches do not provide a physically realistic description of the JJ's irreversible dynamics.  We propose instead to treat the JJ explicitly as an open system.  In the last part of this section we discuss how the laws of thermodynamics require that there be a positive feedback, allowing the JJ to run as an engine.

\subsection{Macroscopic wave function} 
\la{sec:MWF}

\begin{figure}[t]
	\subfigure[]{\includegraphics[width=0.3 \textwidth]{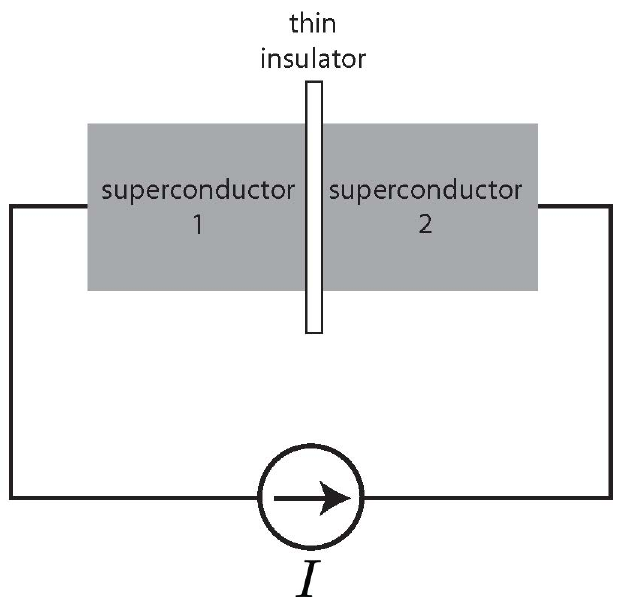}} \hskip 2.5 cm
	\subfigure[]{\includegraphics[width=0.25 \textwidth]{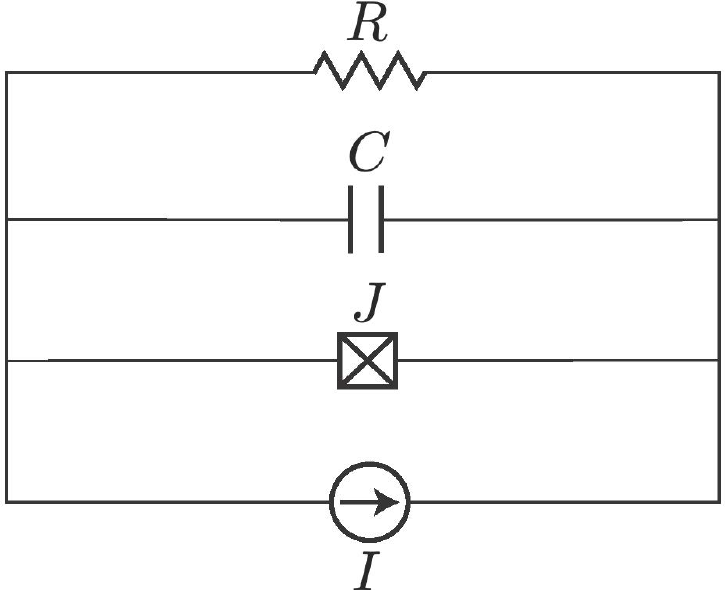}}
\caption{(a) Josephson junction (JJ), composed of two superconducting electrodes separated by a thin insulator.  An external current $I$ is applied to the JJ by an external source.  (b) In the RCSJ model, this JJ is treated as an idealized junction $J$ obeying the current-phase relation of \Eq{eq:DC}, in parallel (``shunt'') with a resistor $R$ and a capacitor $C$.\la{fig:junction}}
\end{figure}

Feynman presented a simple model for JJ in \cite{Feynman}, based on a two-component MWF for the Cooper-pair condensate:
\be
\Psi = \begin{pmatrix}
\psi_1  \\
\psi_2 \\
\end{pmatrix} = 
\begin{pmatrix}
\sqrt{n_1} e^{i\phi_1} \\
\sqrt{n_2} e^{i\phi_2} \\
\end{pmatrix} ,
\la{eq:JJwave}
\ee
where $n_{1,2}$ are Cooper-pair numbers and $\phi_{1,2}$ are phases corresponding to the left-hand and right-hand electrode of the JJ, respectively (see \Fig{fig:junction}). The dynamics is given by the Schr\"odinger equation
\be
\frac{d}{dt}\Psi(t) = -\frac{i}{\hbar} H \Psi(t)
\la{eq:JJSeq}
\ee
with Hamiltonian
\be
H =\hbar
\begin{pmatrix}
U/2 & K \\
K & - U/2 \\
\end{pmatrix} ,
\la{eq:hamJJ}
\ee
where $\hbar U$ is the energy difference between a Cooper pair located at the right electrode and one located at the left electrode. The value of $U$ is related to the voltage bias $V$ across the JJ by the formula 
\be
U= \frac{2e V}{\hbar} ,
\la{eq:voltage}
\ee
while $K$ is the rate of tunneling of Cooper pairs across the barrier between the two electrodes.

In this treatment, the tunneling current is obtained from the unitary evolution according to Eqs.\ \eqref{eq:JJSeq}, \eqref{eq:hamJJ}, and \eqref{eq:voltage}, but the effect of that current on the charge of the electrodes is ignored.  As Feynman explained, this is because the electrodes are connected to the terminals of a battery, which supplies the necessary current to keep the density of Cooper pairs in the superconducting electrodes approximately constant, so that the net charge density of the corresponding material (including the fixed positive charge density of the underlying atomic lattice) remains close to zero \cite{Feynman}.  This coupling to the terminals of the battery makes the JJ an open system.  Describing the JJ using closed-system equations of motion (plus the {\it ad hoc} condition of constant Cooper-pair densities in the electrodes)  obscures the irreversible dynamics by which the JJ can output work in the form of a self-oscillating \cite{SO} electrical dipole that generates non-thermal radiation.

\subsection{RCSJ model} 
\la{sec:RCSJ}

Let $I$ be the external current applied to the JJ, as shown in \Fig{fig:junction}.  In the RCSJ model we have
\be
I = I_c \sin \phi + C \dot V + \frac V R ,
\la{eq:Kirch}
\ee
where the first term in the right-hand side of \Eq{eq:Kirch} corresponds to the Josephson current of \Eq{eq:DC}.  This current is taken to run in parallel (``shunt'') with a current $C \dot V$ through capacitor $C$, and also to a current $V/R$ through resistor $R$; see, e.g., \cite{Tinkham, Strogatz-JJ}.  By using \Eq{eq:dphi} to eliminate $V$ in \Eq{eq:Kirch}, we get the equation of motion
\be
\ddot \phi + \frac{\dot \phi}{R C} + \frac{2 e I_c}{\hbar C} \sin \phi = \frac{2 e I}{\hbar C} ~.
\la{eq:WBeom}
\ee
In the $R \to \infty$ limit, \Eq{eq:WBeom} can be obtained from the Hamiltonian
\be
H(q,\phi) = \frac{e}{\hbar C} \cdot q^2 - I \phi - I_c \cos \phi ,
\la{eq:WBH}
\ee
where the charge $q$ separated across the terminals of the junction is taken as a canonically conjugate variable to the phase $\phi$.  However, \Eq{eq:WBH} is problematic in that the angle $\phi$ must be taken as a real number (i.e., a non-compact variable) with a ``tilted washboard potential''
\be
U(\phi) = - I \phi - I_c \cos \phi
\la{eq:washboard}
\ee
that is unbounded from below.

In this article  we will pursue a qualitatively different approach.  By treating the JJ as an open system, we obtain a master equation that naturally incorporates a thermodynamic irreversibility in the form of the coupling of the Cooper pairs to two electronic baths that are out of equilibrium with each other.  We will see that such a treatment incorporates a resistance $R$ for the JJ that is not related to the phase-slip phenomena often invoked in the literature (see, e.g., \cite{Halperin}).  In our model, the \hbox{DC $\to$ AC} dynamics of the JJ can be naturally understood as the irreversible operation of an engine, powered by the thermodynamic disequilibrium between the electrons in the two terminals of the external battery.

\subsection{Feedback in engines}
\la{sec:feedback}

We use the word {\it engine} to denote an open system that can generate positive work over the course of a cycle, at the expense of an external disequilibrium.  In the case of a {\it heat engine}, this disequilibrium is a difference of temperatures between two external baths, which causes heat to pass through the system.  The disequilibrium may also be a difference of chemical potentials between the baths, causing matter (electrons in the case of the JJ) to pass through the open system.  In addition to the external disequilibrium, an engine must exhibit {\it feedback} between the state of a macroscopic ``tool'' (such as a piston or turbine) acted upon by the engine's working substance, and the coupling of that working substance to the external baths.  It is this feedback that causes an active force that does positive work on the tool over a complete cycle of the engine \cite{engines}.

The laws of thermodynamics can helps us to understand better the general nature of this feedback.  Consider an engine with a homogenous working substance that has internal energy $U$, quantity of matter $N$, and chemical potential $\mu$.  By the first law of thermodynamics, an infinitesimal transformation of this working substance gives a change of the internal energy
\be
d U = \delta Q - \delta W + \mu \, d N ,
\la{eq:1stlaw}
\ee
where $\delta Q$ is the infinitesimal heat absorbed by the substance from its surroundings, and $\delta W$ is the infinitesimal work that the substance does on the tool (the symbol $d$ is used for exact differentials, while $\delta$ is used for inexact differentials).  Some authors refer to the $\mu \, d N$ term as ``chemical work'', but here we want to stress the distinction between this term and the actual {\it mechanical} work $\delta W$ that is associated with a force acting on a {\it directional} and macroscopic degree of freedom, such as the position of a piston.  Although the chemical contribution $\mu \, d N$ carries no entropy, its {\it cyclical} conversion into mechanical work is non-trivial: it requires feedback and dissipation, leading to an active force capable of driving the tool; see \cite{engines} and references therein.

Over a full cycle the substance returns to its initial state, so that the change to $U$ vanishes:
\be
\Delta U = \oint \left( \delta Q - \delta W + \mu \, d N \right) = 0 .
\la{eq:conservation}
\ee
The net work is, therefore,
\be
W = \oint \delta W = \oint ( \delta Q + \mu \, d N) .
\la{eq:work}
\ee
By Clausius's theorem, the entropy produced by the engine over one cycle is
\be
\Sigma = - \oint \frac{\delta Q}{T} = - \oint \frac{\delta Q}{\bar T \left ( 1 + T_d / \bar T \right)} \geq 0 ~,
\la{eq:Clausius}
\ee
where $T$ is the temperature of the source from which the working substance receives heat $\delta Q$ at each instant during the cycle, $\bar T$ is the mean value of the temperature over the cycle's full period, and \hbox{$T_d \equiv T - \bar T$}.  Let $\mu$ be the instantaneous chemical potential and $\bar \mu$ its mean, with \hbox{$\mu_d \equiv \mu - \bar \mu$}.  Combining \Eq{eq:work} and \Eq{eq:Clausius} we obtain
\be
W \leq \oint \delta Q \left( 1 - \frac{1}{1 + T_d / \bar T} \right) + \oint \mu \, d N = \oint \frac{\delta Q \cdot T_d}{\bar T + T_d} + \oint \mu_d \, d N ~.
\la{eq:Rayleigh}
\ee
This bound on $W$ is achieved by a reversible cycle ($\Sigma = 0$).  The result of \Eq{eq:Rayleigh} may be generalized to inhomogenous temperatures and chemical potentials by integrating over the maximum work that each part of the working substance may perform.

The result of \Eq{eq:Rayleigh} tells us that for a heat engine to generate positive work ($W > 0$), its working substance must be at a higher temperature ($T_d > 0$) during the part of the cycle during which it is more strongly coupled to the hotter bath (from which it takes heat $\delta Q > 0$).  Conversely, the working substance must be at a lower temperature ($T_d < 0$) when it is more coupled to the colder bath (into which it rejects heat $\delta Q < 0$).  Thus, in any heat engine there must be positive feedback between the working substance's instantaneous $T_d$ and its coupling to the external baths (which gives the instantaneous $\delta Q$) \cite{engines}.  In \cite{engines} this general result is called the ``Rayleigh-Eddington criterion''.\footnote{The first published reference where this result was called the ``Rayleigh-Eddington criterion'' was \cite{solar}. Equation \eqref{eq:Rayleigh} for the particular case $\mu_d = 0$ was written by Eddington in \cite{Eddington1, Eddington2}, in the context of his theory of Cepheid variable stars, which exhibit self-oscillation in their magnitude (i.e., brightness) and volume.  Eddington was apparently unaware that an equivalent criterion had already been formulated qualitatively by Rayleigh in \cite{Rayleigh1, Rayleigh2}, in the context of thermoacoustic oscillations.  In mechanical engineering, that earlier result is commonly referred to as the ``Rayleigh criterion'' for thermoacoustic instabilities (see, e.g., \cite{thermoacoustics}).}

It is also evident from \Eq{eq:Rayleigh} that a heat engine works most efficiently if heat is absorbed by the working substance {\it only} at the maximum $T_d$, while heat is rejected {\it only} at the minimum $T_d$.  In that case, \Eq{eq:Rayleigh} gives the Carnot bound.  If $T_d = 0$ (as in the case of the JJ, where the Cooper pairs are always at the ambient temperature), \Eq{eq:Rayleigh} reduces to
\be
W \leq \oint \mu_d \, dN .
\la{eq:chem-eng}
\ee
Note that \Eq{eq:chem-eng} applies even if the working substance exchanges heat with its surroundings, as long as the temperature is fixed.  Thus, we see that in a {\it chemical engine} (i.e., in an engine which runs on a disequilibrium of the external chemical potentials, rather than on a thermal disequilibrium) there must be a positive feedback such that, during the cycle of operation, matter enters the engine's working substance at a higher chemical potential ($\mu_d > 0$ when $dN > 0$) and exits at a lower chemical potential ($\mu_d < 0$ when $dN < 0$).  The efficiency of a chemical engine will always be less than unity, even without dissipation, because matter must exit at some non-zero potential $\mu$.  But there is no Carnot bound in this case nor (as far as we know!) any other universal bound independent of engine design.

In the models for the ``putt-putt'' steam engine presented in \cite{engines, solar}, as well as in the ``electron shuttle'' and the ``Quincke rotor'' described in \cite{engines} the engine's dynamics is formulated in terms of two coupled ordinary differential equations: a mechanical equation of motion (second order in time) for a macroscopic degree of freedom (the ``tool''), and a kinetic equation (first order in time) for the amount of matter in the working substance.  The kinetic equation is explicitly irreversible (i.e., non-invariant under $t \to -t$).  A feedback appears between those two equations: the working substance exerts a force on the tool, while the tool modulates the coupling of the working substance to the external baths (and therefore the rate of change of the corresponding amount of matter).  This feedback is what makes the force on the tool {\it active}, i.e., a non-conservative force that does positive work ($W > 0$) over a cycle of the tool's motion.  Milburn underlines that this kind of dynamical system (which he applies to the electronic Schmitt trigger in sec.\ 2.3 of \cite{Milburn}) is not Hamiltonian.

The relation between voltage $V$ and phase $\phi$ in the RCSJ model is such that the voltage can be eliminated to give an equation of motion for $\phi$ alone [\Eq{eq:WBeom}].  The absence of a stable equilibrium value for the angle $\phi$ (which is physically defined modulo $2\pi$) requires reinterpreting $\phi$ as a non-compact variable with a potential $U(\phi)$ that is unbounded from below, as in \Eq{eq:washboard}.  Our model of the JJ as an engine is qualitatively different.  Instead of just the phase $\phi$, we will work with the macroscopic coherence $z$ (a complex number whose argument corresponds to $\phi$). The equation of motion for $z$ depends on the separation of charge between the two electrodes (or, equivalently, on the instantaneous voltage jump across the junction).  A feedback appears in that the rate of change of this separated charge depends on $z$ because of the tunneling between electrodes.  The dependence between $z$ and $V$ is such that the dynamics cannot be reduced to an equation of motion for $z$ alone.

When an external current source is applied to the JJ, the feedback can be positive, maintaining $|z|$ against the damping associated with the decay of Cooper pairs in the electrodes into electrons in the external baths.  This positive feedback sustains the rotation of the phase $\phi$.  Thus, in our model the JJ is considered as an engine in which the Cooper pairs in the electrodes constitute the working substance, while the JJ's macroscopic coherence $z$ acts as the tool.  The Rayleigh-Eddington criterion of \Eq{eq:chem-eng} is then met, because more Cooper pairs are formed in the electrode at the higher voltage, while more Cooper pairs are destroyed in the electrode at the lower voltage.  The resulting engine dynamics can also produce self-oscillation of the JJ's electric dipole, thus explaining how the device produces monochromatic radiation.

We will consider the precise dynamics of these various processes in the rest of this paper.  However, it will be useful to keep in mind how the detailed model fits into this simple thermodynamic picture.  At the most basic level, the point is that electrons come from the bath at higher potential, pass through the JJ (in the form of Cooper pairs), and end up in the bath at the lower potential.  This flow sustains the JJ's macroscopic coherence $z$ and turns its phase $\phi$.  This turning phase is a form of {\it work}, which can be used to produce non-thermal radiation.  It may also be used to pump a current in another circuit, as has been shown experimentally in \cite{JJ-pump}.  This work extraction is possible because of a positive feedback between $z$ and the differential rates at which Cooper pairs are produced or destroyed in the electrodes.

In previous theoretical treatments of microscopic engines (in addition to the references already cited see also, e.g., \cite{adiabatic-motors, QT, stochastic-shuttle}) the tool degree of freedom is effectively classical.  Our model of the JJ is new because the tool is the coherence $z$, which is a macroscopic but essentially quantum object.  We therefore expect, even beyond the specific problem of understanding the Josephson effect as an irreversibly process, that the model that we present here may open new perspectives in the theory of quantum thermodynamics.

\section{Irreversible dynamics}
\la{sec:open}

Our model is an extension of the model proposed by Feynman, in which the MWF will be replaced by the {\it macroscopic density matrix} (MDM), which is the appropriate mathematical object to describe the dynamics of the JJ as an open system.\footnote{In some other contexts, this same object is referred to as a ``one-body reduced density matrix''.}  The coupling to baths of electrons gives a non-unitary evolution of this MDM that can be expressed in terms of a Markovian master equation (MME).  From this MME we arrive at a formulation of the dynamics of the JJ in terms of coupled first-order ordinary differential equations for the time-dependence of the JJ's voltage $V$ and its coherence $z$.

\subsection{Markovian master equation}
\la{sec:MME}

Let us begin by considering a single superconducting electrode, coupled to an external bath of electrons. The condensate of Cooper pairs occupies the ground state of a certain effective Hamiltonian, which can be accounted for using a single quantum harmonic oscillator with the operator $a^\dagger a$ for the number of Cooper pairs. This number can be increased by the fusion of two electrons from the bath into a Cooper pair, or decreased by the reverse process of Cooper pair decomposition. The effective Hamiltonian describing such processes takes the form
\be
H = E_0 a^\dagger a + \sum_k \epsilon_k  b_k^\dagger b_k  + \sum_{k,k'} \left( g_{kk'} a b_k^\dagger b_{k'}^\dagger + \bar g_{kk'} a^\dagger b_{k'} b_k \right) 
\la{eq:Htot}
\ee
where $b_k^\dagger$ and $b_k$ are, respectively, creation and annihilation operators for electrons in the bath.  The interaction Hamiltonian of \Eq{eq:Htot} corresponds, when restricted to on-shell processes, to the well known Andreev scattering, given by the coupling between electrons and holes in the normal metal and Cooper pairs in the superconductor \cite{Andreev}.  On the physics of Andreev scattering, see also \cite{Andreev-rev} and references thererin.  The results of this paper will not depend on the detailed form of the $g_{kk'}$ form factors in \Eq{eq:Htot}.

Under the standard assumption of separation of time scales of the system (Cooper pairs) and baths (electrons), one obtains the following Markovian master equation for the reduced density matrix of Cooper pairs (harmonic oscillator):
\be
\frac{d}{dt} \rho = - \frac i \hbar E \left[a^\dagger a, \rho \right] + \frac 1 2 \gd \left( \left[ a , \rho a^\dagger \right] + \left[ a \rho, a^\dagger \right] \right) + \frac 1 2 \gu \left( \left[ a^\dagger, \rho a \right] + \left[ a^\dagger \rho, a \right] \right) 
\la{eq:MME}
\ee
where $E$ is the renormalized value of ground state energy and $ \gd > \gu$ are decomposition and creation rates, respectively. The explicit formulas for those parameters are not relevant at the moment and can be found in the literature (see, e.g., \cite{AL}).  Using \Eq{eq:MME} one can find the exact kinetic equation for the number of Cooper pairs $n(t) = \mathrm{Tr}[\rho(t) a^{\dagger}a]$:
\be
\dot n = - \gamma (n - \bar n) , \quad \gamma \equiv \gd - \gu, \quad \bar n \equiv \frac \gu \gamma .
\la{eq:kinetic}
\ee

The Markovian approximation is justified in this case because of the separation of time scales between the slow dynamics of the system composed of Cooper pairs in the electrodes and the fast internal dynamics of the electrons in the leads.  For the Cooper pairs, the relevant time scales are the Josephson frequency of \Eq{eq:dphi}, the tunneling rate $K$, and the damping rate $\gamma$ (we will define these carefully in \Sec{sec:model}). In a common experimental setup, all of these are $\sim 10^{-12}$ s; see, e.g., \cite{Febvre}.  Meanwhile, the time scale for the internal dynamics of the leads can be estimated from the Drude model of electrical conduction to be $\sim 10^{-14}$ to $10^{-15}$ s.  There is, therefore, at least two orders of magnitude of separation between the respective time scales of the system and the baths for the physical setups that we are interested in describing.

\subsection{Josephson Junction model}
\la{sec:model}

In order to describe the JJ, we must extend the model above so as to describe two superconducting electrodes separated by a potential barrier, each of them coupled to its own electron reservoir.  Consider, therefore, two Cooper-pair modes $1$ and $2$ corresponding to the two electrodes in \Fig{fig:junction}.  The system is then equivalent to two quantum harmonic oscillators, with reduced dynamics given by an extension of \Eq{eq:MME}, to wit:
\be
\frac{d}{dt} \rho = - \frac i \hbar [ H_{12}, \rho]  + \La_1 \rho + \La_2 \rho .
\la{eq:MME1}
\ee
The Hamiltonian $H_{12}$ takes the form
\be
     H_{12} = 
     E_1 a_1^\dagger a_1 + E_2 a_2^\dagger a_2
     + \hbar K \left( a_1^\dag a_2 + a_2^\dag a_1 \right),
\la{eq:Ham2}
\ee
where $E_{1,2}$ are the energies for each of the two electrodes, while $K$ is the tunneling rate between the electrodes, which can be taken positive by absorbing the phase. Note, that we consider the so-called {\it local} master equation \cite{LevyKosloff}.  This approach is justified when the perturbation is small compared to the energy scale of the free Hamiltonian \cite{Brunner-local,Adesso-local}.  In our case, we treat the tunneling rate $K$ as a small perturbation compared to the electrode energies $E_{1,2}$ in \Eq{eq:Ham2}, which is consistent with usual approach in theory of superconductors.

The dissipative generators $\La_{1,2}$ in \Eq{eq:MME1} correspond to the coupling of each of the two electrodes of the JJ to the electronic bath with which that electrode is in physical contact.  These generators have the same structure as in \Eq{eq:MME}, with
\be
\La_j \rho =  \frac 1 2 \gamma^{(j)}_{\downarrow} \left( \left[ a_j , \rho a_j^\dagger \right] + \left[ a_j \rho , a_j^\dagger \right] \right) + \frac 1 2 \gamma^{(j)}_{\uparrow} \left( \left[ a_j^\dagger , \rho a_j \right] + \left[ a_j^\dagger \rho , a_j \right] \right) .
\la{eq:MMEj}
\ee

In order to obtain the generalization of \Eq{eq:kinetic} corresponding to the two electrodes of the JJ, we need to define a single-particle density matrix.  By analogy with the macroscopic wave function of \Eq{eq:JJwave}, we will call this new object the \emph{macroscopic density matrix} (MDM). In this case, the MDM is a $2 \times 2$ positively defined matrix $\sigma$, given by the following two-point correlations:
\be
\sigma \equiv [\sigma_{jk}] , \quad \sigma_{jk} = {\rm Tr}(\rho a_j^\dagger a_k), \quad j,k = 1,2 .
\la{eq:sigma}
\ee
A useful parametrization of MDM is given by the populations $n_j = {\rm Tr}(\rho a_j^\dagger a_j)$ and coherence  $z = {\rm Tr}(\rho a_1^\dagger a_2)$:
\be
\sigma=
\begin{pmatrix}
n_1 & z \\
z^\ast & n_2 \\
\end{pmatrix}, \quad |z|^2 \leq n_1 n_2 .
\la{eq:MDMapp}
\ee
Notice that, for the pure state corresponding to \Eq{eq:JJwave},
\be
\sigma = |\Psi\rangle\langle \Psi | \quad \Rightarrow \quad z = \sqrt{n_1 n_2} e^{i\phi}, \quad \quad \phi= \phi_1 - \phi_2.
\la{eq:MDMpure}
\ee
Eqs.\ \eqref{eq:MME1} -- \eqref{eq:MDMapp} translate into the following evolution equations for the MDM:
\bea
\dot  n_1 &=&- \gamma^{(1)} (n_1 - \bar n_1)  -  i K (z - z^\ast)  \nl
\dot n_2 &=& - \gamma^{(2)} (n_2 - \bar n_2)  +  i K (z - z^\ast)  \la{eq:MDMeqapp} \\
\dot z &=& \left[ \frac{i}{\hbar} U  -  \frac 1 2 (\gamma^{(1)} +\gamma^{(2)}) \right] z  - i K (n_1 - n_2) . \nn
\eea
where $U = E_1 - E_2$, $\gamma^{(j)} \equiv \gamma^{(j)}_{\downarrow}- \gamma^{(j)}_{\uparrow}$, and $\bar n_j \equiv \gamma^{(j)}_\uparrow /\gamma^{(j)}$ for $j=1,2$.
 
Let us assume that the relaxation rates for both electrodes are equal, i.e.,
\be
\gamma^{(1)} = \gamma^{(2)} = \gamma .
\la{eq:gamma}
\ee
Then, introducing new variables for the sum and difference of the expected number of Cooper pairs on each of the JJ's electrodes,
\be
N = n_1 + n_2 , \quad n = n_1 - n_2 ,
\la{eq:variablesNn}
\ee
we rewrite \Eq{eq:MDMeqapp} as
\begin{align}
\dot N &= - \gamma (N - \bar N) , \quad \bar N = \bar n_1 + \bar n_2, \la{eq:MDM1-N} \\
\dot n &= - \gamma(n - \bar n)  - i 2K (z - z^\ast), \quad \bar n = \bar n_1 - \bar n_2, \la{eq:MDM1-n} \\
\dot z & = \left( \frac{i}{\hbar} U  - \gamma \right) z  -  i K n .\la{eq:MDM1-z}
\end{align}
This implies that the average total number of Cooper pairs $N$ relaxes, independently of all the other dynamical variables, to its stationary value ($N = \bar N$), which must correspond physically to an electrically neutral JJ.  This justifies our having taken equal relaxation rates for both electrodes in \Eq{eq:gamma}.

The variable $n$ describes an internal local disequilibrium that causes an electrical double layer to form at the junction, with excess charge $Q = e n$ and associated voltage jump $V$. Accordingly, we introduce the capacitance of the junction $C$, such that the voltage is given by the formula:
\be
V = \frac Q C = - \frac{e n}{C}.
\la{eq:V}
\ee
We can interpret the energy difference $U$ between the two electrodes as the work needed to transfer a single Cooper pair between the electrodes, against the external voltage.  Thus:
\be
U = - 2 e V = \frac{2 e^2 n}{C}.
\la{eq:energy_difference}
\ee
Note that, by making the voltage across the junction depend on the difference $n$ of the numbers of Cooper pairs, we have introduced a {\it mean-field approximation}.  This is justified as long as $n_1, n_2 \gg |n|$.  The tunneling rate $K$ will also depend on $n$, but that dependence ---unlike the dependence of $V$ on $n$--- is not essential to the JJ dynamics that we will describe here.  For reasons of simplicity we will take $K$ as constant.
 
Combining all of the above definitions, we get a dynamical system characterized by the coupled differential equations
\bea
\dot V &=& - \gamma V + i \frac{2Ke}{C} \left(z - z^\ast \right) +  \frac{I}{C}, \la{eq:MDM2-V} \\
\dot z &=& - \left( i \frac{2eV}{\hbar} + \gamma \right) z  +  i \frac{KC}{e} V, \la{eq:MDM2-z} 
\eea
where we have put: 
\be
I = - e \gamma \bar n  = - e \left( \gamma^{(1)}_\uparrow - \gamma^{(2)}_\uparrow \right) .
\la{eq:parameterI}
\ee
Finally, we introduce a resistance parameter $R$ such that
\be
\gamma = \frac{1}{RC} ,
\la{eq:resistance}
\ee
in order to facilitate comparison between our results and those of the RCSJ model.  Using this, we may rewrite \Eq{eq:MDM2-V} as
\be
I = C \dot V + \frac V R + I_J, \quad \hbox{where} \quad I_J = 4 e K \, \Im{z} = 4 e K |z| \sin \phi ,
\la{eq:Kirch2}
\ee
where the last term can also be defined, in terms of the Hamiltonian dynamics, as: 
\be
I_J= q \dot n_1 =
\frac{i q}{\hbar} \Tr \left( [H_{12},  a_1^\dagger a_1] \rho \right) 
= - 2 q K |z| \sin \phi,
\la{eq:josephson_current}
\ee
with $q=-2e$ the charge of a Cooper pair.

Comparing \Eq{eq:Kirch2} to \Eq{eq:Kirch} of the RCSJ model, we interpret the quantity $I$ defined by \Eq{eq:parameterI} as the fixed current that the external source provides to the JJ.\footnote{A full treatment of the JJ's dynamics would require us to include in the Hamiltonian the electrostatic interaction between the Cooper pairs and also with the background of positive ions in the electrodes, as Feynman suggests in \cite{Feynman}.  That interaction makes it energetically costly to vary the local density of Cooper pairs from its equilibrium value.  By making the relaxation rates equal [\Eq{eq:gamma}] (which allows the JJ to relax to  state with net zero electric charge, independently of $n$ and $z$) and by taking the $I$ defined by \Eq{eq:parameterI} to be equal to the fixed current provided by the JJ's coupling to the external baths (an assumption justified by the fact that form of the dependence on $I$ of \Eq{eq:MDM2-V} corresponds to an electrostatic capacitor connected to a current source), the main consequences of this interaction are incorporated into our dynamical equations without having to treat it explicitly.}  We also see that the term $V /R$ in \Eq{eq:Kirch2} results from the finite lifetimes of the Cooper pairs (which decay into electrons in the baths), while the tunneling of Cooper pairs contributes a current \hbox{$4 e K |z| \sin \phi$}.  For simplicity, we neglect the contribution to the JJ current from quasiparticles. Note that $|z|$ is not fixed, because the $n_{1,2}$ of Eqs.\ \eqref{eq:sigma} and \eqref{eq:MDMapp} are dynamical variables.  This is a key difference between our model and that of RCSJ, in which the tunneling current was taken to be $I_c \sin \phi$, for fixed $I_c$.  Note also that \Eq{eq:MDM2-z} implies that the standard relation between voltage and phase as given in \Eq{eq:dphi} and assumed by the RCSJ model, is not exact.  Therefore \Eq{eq:AC} for the AC tunneling current (a quantity that has never been measured directly in experiments), is not applicable to our model.

We see now how the feedback that we discussed in general thermodynamic terms in \Sec{sec:feedback} appears mathematically in our dynamical model: \Eq{eq:MDM2-V} for the real-valued $V$ depends on $z$ via the tunneling term proportional to $K \, \Im{(z)}$.  Meanwhile, \Eq{eq:MDM2-z} for the complex-valued $z$ depends on $V$, which controls the resonant frequency $2eV/\hbar$.  It is therefore plausible that, in some regime, this non-Hamiltonian dynamical system will self-oscillate: An oscillation of $\Im{(z)}$ forces $V$ in in \Eq{eq:MDM2-V} into an oscillation with the same frequency.  Meanwhile, an oscillation of $V$ about a non-zero average value can excite $z$ via parametric resonance in \Eq{eq:MDM2-z}.  Thus we expect that the feedback between $V$ and $z$ may, in certain regimes, lead to a self-oscillation.  We will investigate this question mathematically in \Sec{sec:SO}.

Note that $I > 0$ in \eq{eq:MDM2-V} is analogous to ``population inversion'' in the theory of a laser, in that it corresponds to $\bar n_1 > \bar n_2$ [see Eqs.\ \eqref{eq:MDM1-n} and \eqref{eq:parameterI}], with the energy of the macroscopic quantum state $|1 \rangle$ greater than that of $|2 \rangle$ by the amount $U$ of \Eq{eq:energy_difference}.  The same is true for $I < 0$, because then $n < 0$ and (by \Eq{eq:energy_difference}) we have that the energy of $|2 \rangle$ is greater than that of $|1 \rangle$.  Thus, it is the external current $I$ that makes the JJ an active system, from which work may be extracted.  The JJ is therefore analogous to a laser in that the ``population inversion'' induced by the external $I$ is what, when combined with a positive feedback involving the coherence $z$ (an effect analogous to stimulated emission), can result in a self-oscillation.\footnote{In this context, it may be worth recalling the lesson of Borenstein and Lamb that the essence of lasing is not intrinsically quantum mechanical \cite{BL-laser}.  Sargent, Scully, and Lamb emphasized the similarity of lasing with classical self-oscillation (which they called ``sustained oscillation'') in \cite{SSL-laser}.}

It will be convenient to transform a set of Eqs. \eqref{eq:MDM2-V} and \eqref{eq:MDM2-z} into dimensionless units. First, we separate the second equation into real and imaginary part of $z$ obtaining
\begin{align}
    & C \dot V + \frac{1}{R} V + I_J = I, \nl
    &\dot I_J +  \frac{1}{RC} I_J +  \left( \frac{2 e}{\hbar} I_S - 4 K^2C \right) V = 0, \\
    &\dot I_S + \frac{1}{RC} I_S - \frac{2 e V}{\hbar} I_J = 0,  \nn
\end{align}
where we have introduced the variable \hbox{$I_S = 4 e K \, \Re{z} = 4 e K |z| \cos \phi$}, so that
\be
z = \frac{1}{4 e K} (I_S + i I_J) .
\la{eq:z-ISJ}
\ee
Let us also introduce the characteristic voltage and current parameters
\be
    \tilde V = \frac{\hbar}{2 e RC}, \quad \tilde I = \frac{\tilde V}{R} ,
\la{eq:charVI}
\ee
and define new dimensionless variables as:
\be \label{eq:dimless_vars}
    \tau = \gamma t, \quad v = V/\tilde V, \quad i_{\rm tot} = I/\tilde I, \quad i_J = I_J/\tilde I, \quad i_S = I_S/\tilde I.
\ee
The equations of motion for the JJ can now be expressed as
\begin{align} \la{eq:dimless_set}
    &\frac{d}{d \tau} v + v + i_J = i_{\rm tot}, \nl
    &\frac{d}{d \tau} i_J +  i_J +  (i_S -4 \a) v = 0, \\
    &\frac{d}{d \tau} i_S +  i_S - i_J v = 0, \nn
\end{align}
where
\be
\a \equiv \frac{K^2}{\gamma^2} .
\la{eq:alpha}
\ee
Our model is therefore characterized by a single parameter $\a$: the ratio of the squares of the tunneling and dissipation rates.

\section{Current-voltage characteristic}
\la{sec:characteristic}

From the dynamical Eqs.\ \eqref{eq:MDM2-V} and \eqref{eq:MDM2-z} we may deduce the stationary current-voltage ($I$-$V$) relation (or ``characteristic'') for the JJ. One can easily solve the fixed point equations ($\dot V = \dot z = 0$) by eliminating $z$ to obtain
\be
I \, R =   V \left[1 + \frac{ 4 K^2}{\left( \frac{2eV}{\hbar} \right)^2 + (RC)^{-2}} \right] .
\la{eq:IVrelation}
\ee
The second term inside the brackets on the right-hand side of \Eq{eq:IVrelation} is a Lorentzian function centered at $V=0$.  It comes from the tunneling current's contribution to the DC component of \Eq{eq:Kirch2}, a contribution that is important only near $V=0$.  For $K \neq 0$, the average power consumed by the JJ is greater than the ohmic $P_{\rm diss} = I V$.  It is this additional power that makes the JJ an active element.  In the dimensionless units introduced before (see Eqs.\ \eqref{eq:charVI}, \eqref{eq:dimless_vars}, and \eqref{eq:dimless_set}), the $I$-$V$ characteristic of \Eq{eq:IVrelation} takes the simple form:
\be
    i_{\rm tot}(v) = v \left( 1 + \frac{4 \a}{1 + v^2} \right) .
\la{eq:i-v_char_dimless}
\ee

\begin{figure}[t]
	\subfigure[]{\includegraphics[width=0.47 \textwidth]{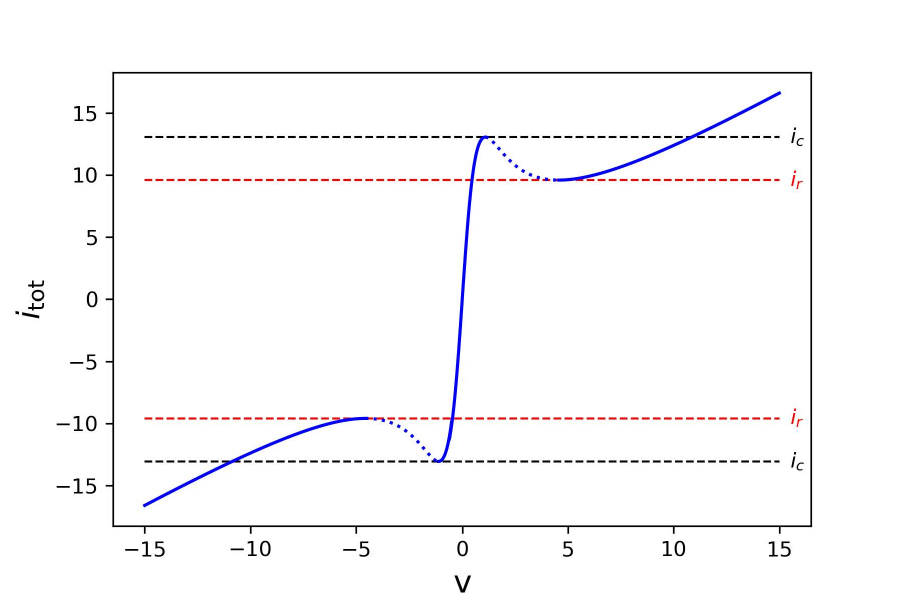}}
	\subfigure[]{\includegraphics[width=0.47 \textwidth]{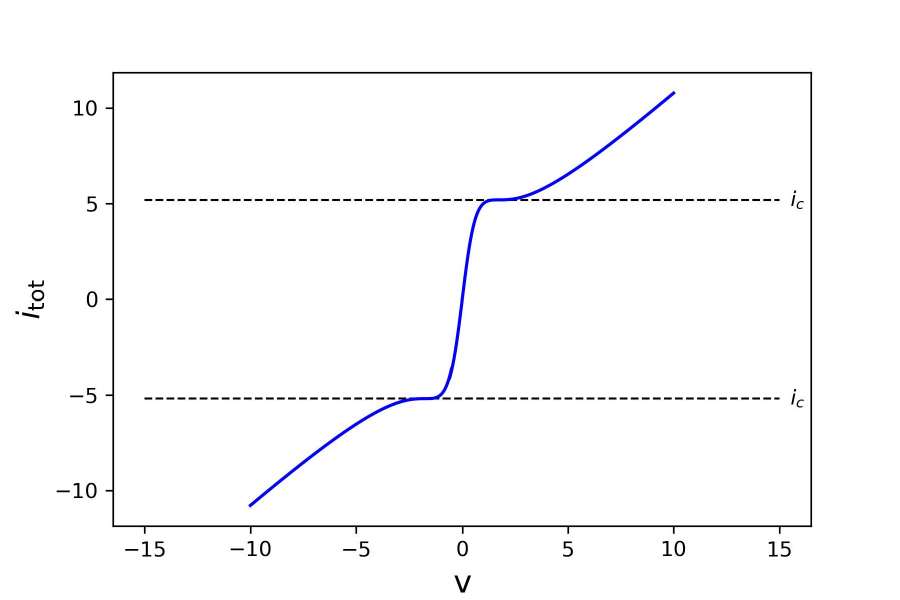}}
\caption{Current-voltage characteristic [\Eq{eq:i-v_char_dimless}], expressed in terms of dimensionless total applied current $i_{\rm tot}$ and voltage $v$ across the junction [for definitions of these dimensionless variables, see \Eq{eq:dimless_vars}]: (a) For $\a = 6$.  The critical current $i_c$ [\Eq{eq:critical}] and the retrapping current $i_r$ [\Eq{eq:retrap}] are marked by the horizontal dashed lines.  The part of the characteristic with negative slope is drawn as dotted rather than solid (see \Sec{sec:stab} for the demonstration that points along this negative slope are unstable equilibria).  (b) For $\a = 2$, in which case the characteristic is monotonically increasing and no hysteresis is observed.\la{fig:IV}}
\end{figure} 

\subsection{Critical and re-trapping currents}
\la{sec:critical}

For $\a > 2$ (i.e., $K > \sqrt 2 \gamma$) the function $i_{\rm tot}(v)$ given in \Eq{eq:i-v_char_dimless} is not monotonically increasing.  The two positive solutions to $i_{\rm tot}'(v_0) = 0$ are
\be
v_{\pm} = \sqrt{2 \a -1 \pm 2 \sqrt{(\a-2) \a}},
\la{eq:OnegR}
\ee
such that a negative differential resistance, i.e., $i_{\rm tot}'(v) < 0$, is observed for voltages $v \in [v_-, v_+]$ or $v \in [-v_+, -v_-]$ (provided that $\a > 2$). The positive local maximum $i_{\rm tot}(v_-)$ we identify with the critical Josephson current:
\be
i_c \equiv i_{\rm tot}(v_-) = (1 + \a + \sqrt{(\a-2)\a}) \sqrt{2\a -1 - 2 \sqrt{\a(\a-2)}}.
\la{eq:critical}
\ee
The positive local minimum $i(v_+)$ (which we will identify with the retrapping Josephson current) is:
\be
i_r \equiv i_{\rm tot}(v_+) = (1+\a - \sqrt{(\a-2)\a}) \sqrt{2\a -1 + 2 \sqrt{\a(\a-2)}}.
\la{eq:retrap}
\ee
Thus, the critical value $\a = 2$ ---for which $i_c = i_r = 3\sqrt{3}$--- discriminates between two regimes: with and without the negative differential resistance of JJ.  In \Sec{sec:stab} we will show that the points along the part of the characteristic curve that have negative differential resistance are unstable, explaining the hysteresis of the underdamped JJ.

Note that \Eq{eq:critical} defines a critical current only for $\a > 2$ (the underdamped regime).  For $\a \leq 2$ (the overdamped regime), a natural definition of the critical current is given by the inflection point: $i_c=i_{tot}(v_0)$ when $i_{tot}''(v_0)=0$.  This corresponds to $v_0 = \sqrt 3$, so that 
\be
    i_c=\sqrt{3}(1+\a), \quad \text{for} \quad \a  \leq2.
\ee
Hence, the expression for critical current for full range of parameter $\a$ is given by 
\begin{align}
    i_c=
    \left\{
    \begin{tabular}{ll}
    	$\sqrt{3}(1+\a)$ & \quad \text{for} $\quad 0 \leq \a \leq 2$ \\
	$(1 + \a + \sqrt{(\a-2) \a}) \sqrt{2 \a - 1- 2 \sqrt{\a(\a-2)}}$ & \quad \text{for} $\quad \a > 2$
    \end{tabular}
    \right.
\la{eq:ic}
\end{align}

These results are illustrated in \Fig{fig:IV}, for two different choices of the parameters $\a$. In \Fig{fig:IV}(a), for $\a > 2$, the characteristic $i_{\rm tot}(v)$ has local maxima and minima, and therefore well-defined values of $i_c$ and $i_r$.  In \Fig{fig:IV}(b) the value $\a \le 2$ the characteristic $i_{\rm tot}(v)$ is monotonically increasing and therefore has no $i_r$.  We will show in \Sec{sec:stab} that the first characteristic exhibits hysteresis, whereas the second does not.

Note that the critical current $i_c$ is given by the expression:
\be
i_c = \frac{I_c}{\tilde I} = \frac{2 e R^2 C I_c}{\hbar} \equiv \beta_c,
\la{eq:ic-beta}
\ee
which agrees with the definition of the Stewart-McCumber parameter $\beta_c$ in the RCSJ model. Thus, our model exhibits hysteresis for $\beta_c \ge 3\sqrt{3}$, whereas in the RCSJ model, hysteresis is usually said to correspond to $\beta_c \gtrsim 1$.

In the limit $\a \to \infty$ (corresponding to a damping rate $\gamma$ much smaller than the tunneling rate $K$), \Eq{eq:ic} implies that $i_c \to 2 \a$.  In that case the physical value of the critical current $I_c$ is
\be
I_c = \frac{\hbar}{2e R^2 C} 2 \a = \frac{\hbar C}{e} K^2 .
\la{eq:Ic0}
\ee
We will use this limiting value of the critical current in \Sec{sec:superMDM}, where we study the efficiency of the JJ as an engine.

\subsection{Stability and hysteresis}
\la{sec:stab}

To determine the stability of fixed points we introduce the small deviations $\delta v$, $\delta i_J$, and $\delta i_S$ of the dynamical variables away from their respective stationary values $v_0, i_{J,0}$, and $ i_{S,0}$, such that
\be
v = v_0 + \delta v , \quad i_J = i_{J,0} + \delta i_J, \quad i_S = i_{S,0} + \delta i_S, \quad \text{where} \quad i_{J,0} = \frac{4 \a v_0}{1+ v_0^2}, \quad i_{S,0} = \frac{4 \a v_0^2}{1+ v_0^2}.
\la{eq:deviations}
\ee
Inserting this into \Eq{eq:dimless_set} yields the linearized equation
\bea
\frac{d}{d\tau}
\begin{pmatrix} \delta v \\  \delta i_J \\ \delta i_S\end{pmatrix} = 
\begin{pmatrix} - 1 & - 1 &  0 \\
\frac{4 \a}{1+v_0^2} & - 1 &  - v_0 \\
\frac{4 \a v_0}{1+v_0^2} &  v_0 & -1 \end{pmatrix}
\begin{pmatrix} \delta v \\  \delta i_J \\ \delta i_S\end{pmatrix} .
\la{eq:MDMlin}
\eea
The eigenvalues of this linear set are therefore the three roots $\{\lambda_i\}$ of the characteristic polynomial
\be
\lambda^3 + 3 \lambda^2 + \left(v_0^2 + \frac{4 \a}{1+ v_0^2} + 3 \right) \lambda +  v_0^2 + 4 \a \ \frac{1- v_0^2}{1+ v_0^2} + 1 .
\la{eq:jacobroots} 
\ee
According to the Routh-Hurwitz criterion, the equilibrium at $\delta v = \delta i_J = \delta i_S = 0$ will be unstable (i.e., at least one root will have $\Re{\lambda_i} > 0$) if and only if
\be
v_0^2 + 4 \a \ \frac{1- v_0^2}{1+ v_0^2} + 1 < 0 .
\la{eq:unstable}
\ee
Note that
\be
v_0^2 + 4 \a \, \frac{1-v_0^2}{1+ v_0^2} + 1 = \left( 1+ v_0^2 \right) \frac{d}{d v} \left[v \left( 1 + \frac{4 \a}{1 + v^2} \right) \right]_{v=v_0}
= \left(1+v_0^2 \right) \, i'_{\rm tot}(v_0) ,
\la{eq:neg-res}
\ee
such that the unstable regime given by condition \eqref{eq:unstable} precisely corresponds with negative differential resistance, $i_{\rm tot}'(v_0) < 0$.

\begin{figure}[t]
	\subfigure[]{\includegraphics[width=0.52 \textwidth]{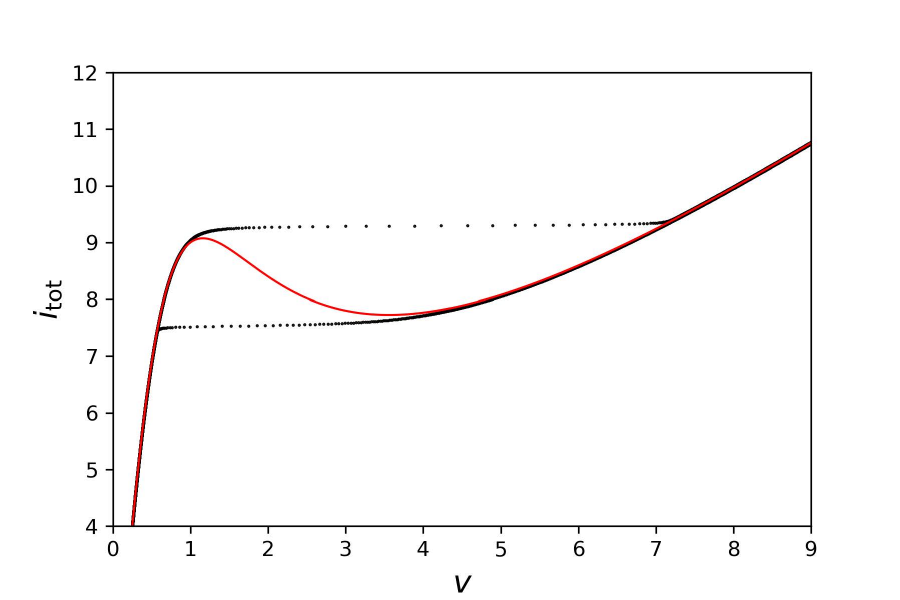}} \\
	\subfigure[]{\includegraphics[width=0.45 \textwidth]{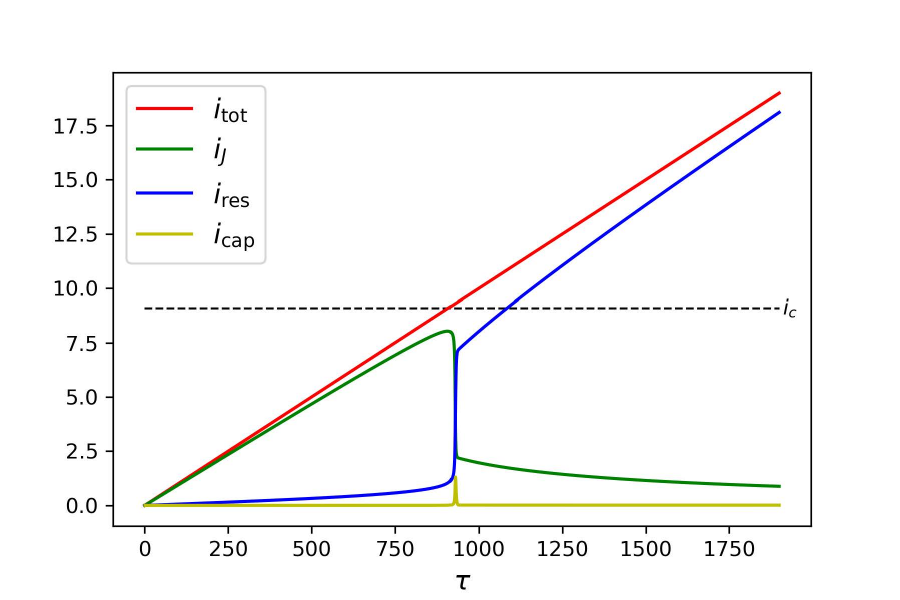}}
 \subfigure[]{\includegraphics[width=0.45 \textwidth]{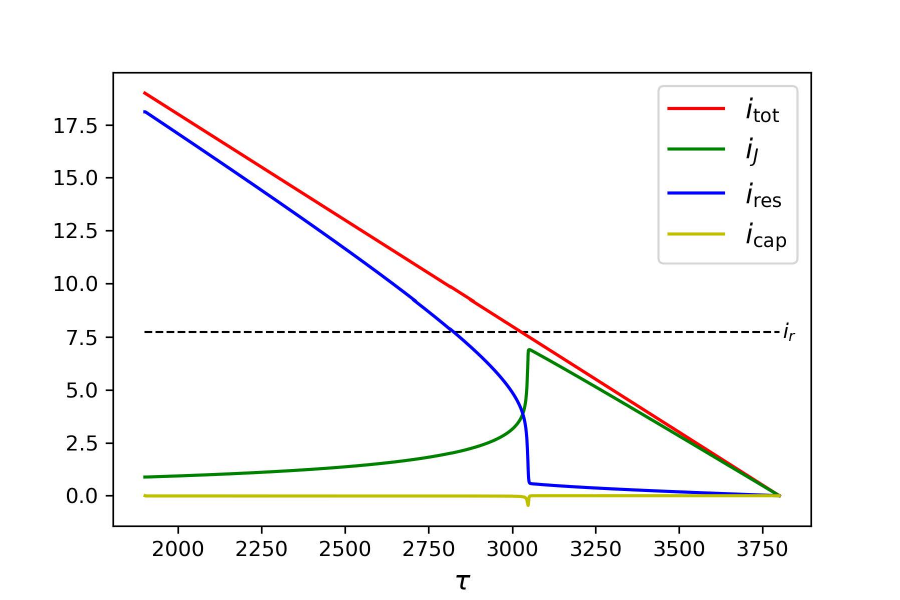}}
\caption{Hysteretic response of the JJ model of \Eq{eq:dimless_set}, with $\a = 4$ and a time-dependent total current $i_{\rm tot}(\tau)$. (a) A hysteresis loop is seen when $i_{\rm tot}$ is first increased and then decreased linearly, with $d i_{\rm tot} / d \tau = \pm 0.01$. The data points obtained by numerical simulation are shown in black.  The solid red curve shows the stationary characteristic of \Eq{eq:i-v_char_dimless}.  (b) Responses of the Josephson (tunneling) current $i_J$, the resistive current $i_{\rm res} = v$, and the capacitive current $i_{\rm cap} = dv/d\tau$ (such that \hbox{$i_J + i_{\rm res} + i_{\rm cap} = i_{\rm tot}$}), during the rising phase ($d i_{\rm tot} / d \tau = 0.01$).  The value of the critical $i_c$ [\Eq{eq:critical}] is marked by a horizontal dashed line.  (c) Responses of $i_J$, $i_{\rm res}$, and $i_{\rm cap}$ during the falling phase ($d i_{\rm tot}/ d \tau = -0.01$).  The value of the retrapping $i_r$ [\Eq{eq:retrap}] is marked by a horizontal dashed line.\la{fig:hyster}}
\end{figure} 

The instability of points along the current-voltage characteristic that have negative slope implies that, for $\a > 2$ (i.e., in the underdamped regime), there will be a hysteresis loop bounded from above by $i_c$ (\Eq{eq:critical}) and from below by $i_r$ (\Eq{eq:retrap}).  If the applied current $i_{\rm tot}$ is initially zero and is then increased slowly, the resulting voltage will follow the nearly vertical segment of the characteristic (see \Fig{fig:IV}(a)) until the  critical current $i_c$ is reached.  Since the part of the characteristic with negative slope is unstable, the current will then stay very nearly constant as the voltage $v$ jumps quickly to the value corresponding to the intersection of the approximately ohmic branch of the characteristic with the horizontal line $i_{\rm tot} = i_c$.  Further increase of the current will move the voltage up along the ohmic branch.  If the voltage is then slowly dialed down, the voltage will decrease smoothly until the value $i_{\rm tot} = i_r$ is reached, whereupon the current will stay almost fixed as the voltage jumps quickly to a value close to zero (corresponding to the intersection of the nearly vertical branch of the characteristic with the horizontal line $i_{\rm tot} = i_r$).

We have verified this hysteretic behavior in numerical integrations of \Eq{eq:dimless_set} with a time-dependent bias current \hbox{$i_{\rm tot} = i_{\rm tot} (\tau)$}.  In particular, we let $i_{\rm tot}$ start at zero, increase linearly up to a value above $i_c$, and then decrease linearly until it returns to zero.  The response of the JJ is shown in \Fig{fig:hyster} for $\a = 4$.  The hysteresis loop appears clearly in \Fig{fig:hyster}(a).

We may also study the responses of the Josephson (i.e., tunneling) current $i_J$, the resistive current $i_{\rm res} = v$, and the capacitive current $i_{\rm cap} = dv/d\tau$, such that $i_J + i_{\rm res} + i_{\rm cap} = i_{\rm tot}$.  In \Fig{fig:hyster}(b) we see a sudden jump in $i_{\rm res}$ shortly after $i_{\rm tot}$ surpasses the critical value $i_c$.  In \Fig{fig:hyster}(c) we see how $i_{\rm res}$ decreases rapidly shortly after $i_{\rm tot}$ falls below the retrapping $i_r$.  These jumps in $i_{\rm res}$ are accompanied by changes in $i_J$ in the opposite direction, as well as by brief spikes in $i_{\rm cap}$.

\subsection{Bifurcation analysis}
\la{sec:bifurcation}

Much like in the ``putt-putt'' engine model of \cite{engines, solar} and in the treatment of the electronic Schmitt trigger in \cite{Milburn}, ours is a non-Hamiltonian dynamical system in which an explicitly irreversible ``kinetic equation'' [\Eq{eq:MDM2-V}] is coupled to a ``mechanical'' equation of motion for an oscillator [\Eq{eq:MDM2-z}].  Positive feedback between these equations can lead to a self-oscillation, which in the models of \cite{engines, solar, Milburn} appears mathematically as a Hopf bifurcation \cite{Strogatz-Hopf}.  It is therefore natural to ask whether something similar happens in our model of the JJ.

Of the three roots of the polynomial in \Eq{eq:jacobroots}, one is real (we label it as $\lambda_0$) and the other two are conjugate:
\be
\lambda_{\pm} = \kappa \pm i \eta , \quad \kappa, \eta \in \mathbb R .
\ee
It can be shown that for $\a > 0$ the real part of the complex roots is non-positive, i.e., $\kappa \le 0$. This implies that a bifurcation appears only for the real $\lambda_0$, which is therefore {\it not} a Hopf bifurcation (i.e., the unstable eigenvalue is not associated with an oscillation frequency $\eta > 0$).  Thus, our model cannot exhibit limit cycles around unstable equilibria, as in the self-oscillating dynamical systems considered in \cite{SO}.

We will return to the question of self-oscillation in \Sec{sec:SO}.  There we will show that our model does exhibit self-oscillation, but that this appears as a {\it hidden attractor} around a linearly stable fixed point, rather than as a limit cycle associated with a Hopf bifurcation \cite{hidden}.  It will therefore require a small but finite perturbation away from the fixed point to get the system to self-oscillate.

\subsection{Interference}
\la{sec:squid}

\begin{figure}[t]
	\includegraphics[width=0.32 \textwidth]{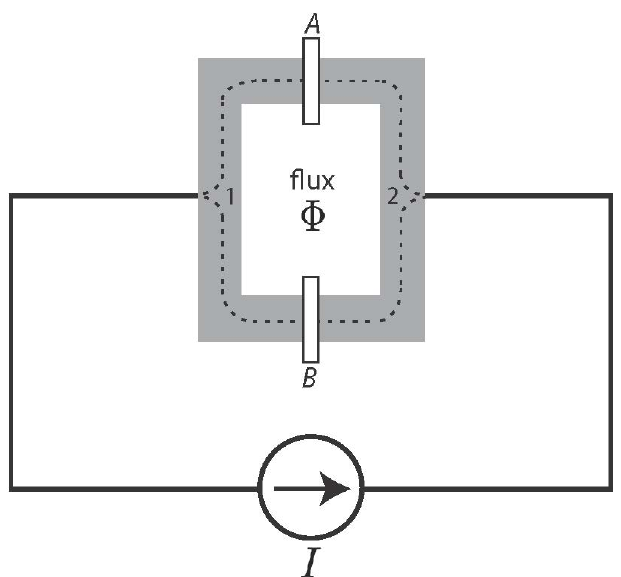}
\caption{Scheme of a superconducting quantum interference device (SQUID).  Current between the two superconducting electrodes (1 and 2) may tunnel either through the thin insulator labelled $A$ or the thin insulator labeled $B$.  An external magnetic flux $\Phi$ is applied through the loop formed by the two dotted paths.\la{fig:squid}}
\end{figure}

Our model can also be applied to interference phenomena.  Consider the superconducting quantum interference device (SQUID) represented in \Fig{fig:squid}.  We modify the tunneling constant $K$ by introducing an external magnetic field (via the vector potential $\bf A$) and we introduce the two tunneling channels through the insulators $a$ and $b$, such that
\be
    K = K_A \exp \left( -i\frac{2e}{\hbar} \int_A \bf A \cdot d\bf s \right) +K_B \exp \left( -i\frac{2e}{\hbar} \int_B \bf A \cdot d\bf s \right) ,
\la{eq:KAB}
\ee
where $K_{A,B}$ are the tunneling rates and $\int_{A,B}$ are the integrals along the paths passing through insulators $A$ and $B$, respectively.  The two terms in the right-hand side of \Eq{eq:KAB} correspond to two possible tunneling paths, associated with the two junctions.

So far we have considered a real tunneling constant.  For complex $K$, the tunneling part of the Hamiltonian \eqref{eq:Ham2} should be rewritten as 
\be
    \hbar  K^\ast a_1^\dagger a_2 + \hbar K a_2^\dagger a_1 .
\ee
We then obtain the following equations:
\begin{align}
\dot n &= - \gamma(n - \bar n) - i 2K^\ast z + i2K z^\ast, \quad \bar n = \bar n_1 - \bar n_2, \la{eq:MDM1-n-complex} \\
\dot z & = \left( \frac{i}{\hbar} U - \gamma \right) z - i K n .\la{eq:MDM1-z-complex}
\end{align}
It follows that the $I$-$V$ characteristics has the same form as in \Eq{eq:i-v_char_dimless}, with $\a = K^2 / \gamma^2$ replaced with $|K|^2 / \gamma^2$. The expressions for the maximal and retrapping currents are also the same as before, with $K$ replaced by $|K|$, where 
\bea
    |K| = \left| K_A \exp \left( -i\frac{2e}{\hbar} \int_A \bf A \cdot d\bf s \right) +K_B \exp \left( -i\frac{2e}{\hbar} \int_B \bf A \cdot d\bf s \right) \right| = \left| K_A +K_B \exp \left( -i\frac{2e}{\hbar} \oint \bf A \cdot d\bf s \right) \right| .
\eea
It we write the magnetic flux through the loop as $\Phi = \oint \bf A \cdot d\bf s$, this becomes
\be
    |K|^2 = |K_A|^2 + |K_B|^2 + 2 |K_A K_B| \cos \left(\frac{2e}{\hbar} \Phi \right),
\ee
where the last term describes a well-known interference effect.  The critical current is almost proportional to $|K|^2$ and therefore also oscillates as a function of the flux $\Phi$.

\subsection{Shapiro steps} 
\la{sec:Shapiro}

Josephson had predicted in \cite{Josephson} that if a JJ were driven by an external oscillating electric field with frequency $\Omega_f$, then the $I$-$V$ characteristic would be modified, with steps appearing at values of the bias voltage $V_0$ such that
\be
\omega_0 \equiv \frac{2 e V_0}{\hbar} = n \Omega_f
\la{eq:Shapiro}
\ee
for $n \in \mathbb{Z}$.  This effect was found experimentally by Shapiro \cite{Shapiro}, and is now used as a metrological standard for the determination of voltages \cite{standard}.

The usual explanation of this effect (see, e.g., \cite{Tinkham}) is that a voltage bias
\be
V = V_0 + V_1 \cos \Omega_f t
\la{eq:VAC}
\ee
causes, by Eqs.\ \eqref{eq:DC} and \eqref{eq:dphi}, a tunneling current
\be
I_J (t) = I_c \sin \left( \phi_0 + \omega_0 t + \frac{2 e V_1}{\hbar \Omega_f} \sin \Omega_f t \right) .
\la{eq:AC-1}
\ee
The sine of the sine can be expanded using Bessel functions, so that
\be
I_J (t) = I_c \sum_{n = -\infty}^\infty (-1)^n J_n \left( \frac{2 e V_1}{\hbar \Omega_f} \right) \sin ( \phi_0 + \omega_0 t - n \Omega_f t) .
\la{eq:AC-2}
\ee
The terms of the expansion in \Eq{eq:AC-2} all average out to zero over long times unless $\omega_0 = n \Omega_f$ exactly, in which case a DC contribution
\be
\bar I_J = (-1)^n J_n \left( \frac{2 e V_1}{\hbar \Omega_f} \right) \sin \phi_0
\la{eq:DC-1}
\ee
would appear.  Although this argument agrees with the observed locations of the Shapiro steps, it fails to account for their widths \cite{Tinkham}.  Various elaborations of the theory have therefore been proposed, based on phenomena such as frequency mixing \cite{Grimes}, resistive feedback \cite{Russer}, or non-linear resonance \cite{Pippard}.  In general, these models assume that the external current $I$ applied to the JJ also oscillates with frequency $\Omega_f$ even though in experiments (including the original one by Shapiro) what is applied to the JJ is an oscillating electric field.

Unlike in the RCSJ model, in our model it is straightforward to account for the effects of this external AC field.  For this, we extend the Hamiltonian of \eqref{eq:Ham2} to include the dipole interaction with the field:
\be
     H_{12} = 
     E_1 a_1^\dagger a_1 + E _2 a_2^\dagger a_2
     + \hbar K \left( a_1^\dag a_2 + a_2^\dag a_1 \right) 
     + e d \varepsilon \cos(\Omega_f t) \left( a_1^\dag a_1 - a_2^\dag a_2 \right). 
     \la{eq:Ham2-Shapiro}
\ee
where $\varepsilon \cos(\Omega_f t)$ is a time-dependent external field with the amplitude $\varepsilon$ and frequency $\Omega_f$, while $e d (a_1^\dag a_1 - a_2^\dag a_2)$ is an electrical dipole moment of JJ.  This gives rise to the following inhomogeneous extension of Eqs.\ \eqref{eq:dimless_set}:
\begin{align} \la{eq:dimless_set_forced}
    &\frac{d}{d \tau} v + v + i_J = i_{\rm tot}, \nl
    &\frac{d}{d \tau} i_J +  i_J +  i_S [v + v_f \cos(\omega_f \tau)] - 4 \a v = 0, \\
    &\frac{d}{d \tau} i_S +  i_S - i_J [v + v_f \cos(\omega_f \tau)] = 0 ,  \nn 
\end{align}
where
\be
    v_f = \frac{d \varepsilon}{\tilde V} \quad \text{and} \quad \omega_f = \frac{\Omega_f}{\gamma}.
\ee

\begin{figure}[t]
	\includegraphics[width=0.55 \textwidth]{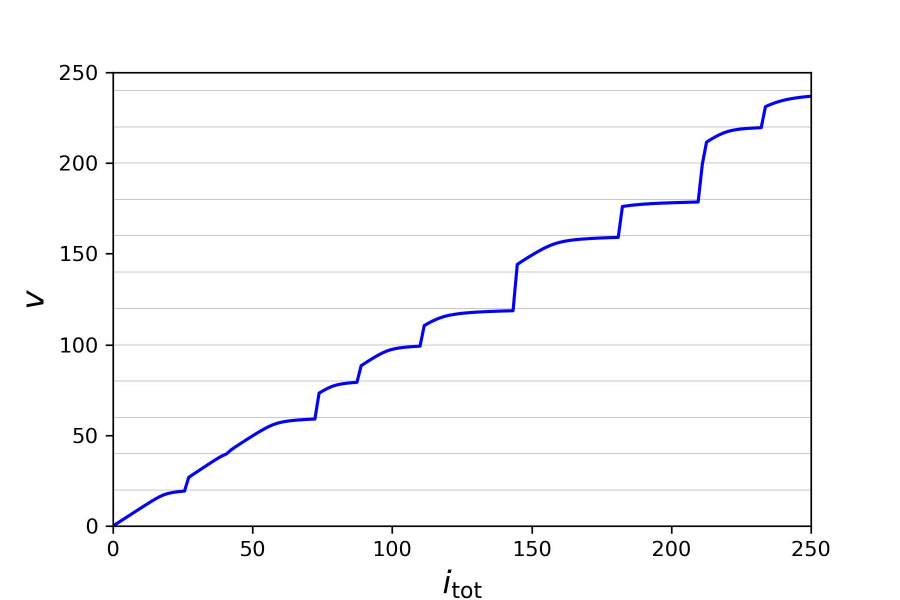}
\caption{Current-voltage characteristic of the Josephson junction in the presence of an oscillating external field with frequency $\omega_f = 20$ and the field voltage $v_f = 300$ for $\a = 3$. The $I$-$V$ characteristics reveals the common Shapiro's staircase structure, where the grey horizontal lines correspond to $v = k \omega_f$ for integer values of $k$.\la{fig:Shapiro}}
\end{figure}

Figure \ref{fig:Shapiro} shows the $I$-$V$ characteristic of the JJ in the presence of an oscillating electric field, with the voltage averaged over many periods of the applied field. The standard Shapiro-step structure is observed, with steps appearing at values given by \Eq{eq:Shapiro}, which in terms of the dimensionless variables introduced above is simply $v = k \omega_f$, for integer $k$.

\section{Self-oscillation}
\la{sec:SO}

We have seen that the simple dynamical model of Eqs.\ \eqref{eq:MDM2-V} and \eqref{eq:MDM2-z} can explain the hysteresis seen in the current-voltage relation for the JJ when the dissipation is small, as well as the frequency locking (``Shapiro steps'') response to an external driving by an applied AC electric field.  Remarkably, the same model exhibits self-oscillations of the voltage and the electrical dipole around stable points in the $I$-$V$ characteristic (i.e., along parts of the characteristic with positive differential resistance).  Even though those fixed points are stable, a finite but small perturbation can move the system out of the basin of attraction of the fixed point, leading to a small self-oscillation with frequency $\Omega = 2 e V / \hbar$, where $V$ is the voltage bias.  This corresponds to a ``hidden attractor'' \cite{hidden}.

In \cite{hidden} the authors draw a distinction between ``self-excited'' oscillations, associated with an unstable fixed point, and ``hidden'' oscillations which are not.  An infinitesimal perturbation is enough to take a self-excited system away from the fixed point and towards the limit cycle or strange attractor.  The lack of such an unstable equilibrium makes the study and simulation of hidden attractors considerably more difficult.  On the other hand, the hidden-attractor oscillations that we consider here for our model of the JJ fall within the scope of Andronov et al.'s conceptualization of a ``self-oscillator'' as an open system that generates and maintains a periodic variation at the expense of an external source of power that has no corresponding periodicity \cite{AVK, SO}.  The presence of hidden attractors in our model of the JJ can explain how the JJ acts as an engine, generating work in the form of non-thermal sound and electromagnetic waves at the frequency $\Omega$ and its lower harmonics.

\subsection{Hidden attractors}
\la{sec:hidden}

The prediction and analytical characterization of hidden attractors is an open problem in the mathematical theory of dynamical systems.  Here we will present a simple argument why the dynamical system described by \Eq{eq:dimless_set} may exhibit hidden limit cycles.  The equations of motion for the system can be expressed as
\bea
&& \dot v + v + \frac{1}{2i}(\zeta - \zeta^*) = i_{\rm tot} , \la{eq:dimless_zeta1} \\
&& \dot \zeta + \zeta + i (\zeta - 4\a) v = 0 . \la{eq:dimless_zeta2}
\eea
where $\zeta$ is a complex variable such that $\Re \zeta = i_S$ and $\Im \zeta = i_J$, while the dot indicates derivation with respect to the dimensionless time variable $\tau$ [see Eqs.\ \eqref{eq:charVI} \eqref{eq:dimless_vars}, and \eqref{eq:dimless_set}].  In terms of the Fourier expansions
\bea
v(\tau) &=& \sum_n v_n e^{-in\omega \tau} , \quad v_{-n} = v_n^* \nl
\zeta (\tau) &=&\sum_n \zeta_n e^{-in\omega \tau},
\la{eq:Fourier}
\eea
and taking $i_{\rm tot}$ = const., Eqs.\ \eqref{eq:dimless_zeta1} and \eqref{eq:dimless_zeta2} become
\begin{align}
&(1 - i n \omega) v_n + \frac{1}{2i} (\zeta_n - \zeta_{-n}^*) - i_{\rm tot} \delta_{n,0} = 0,\la{eq:Fourier1} \\
&(1 - i n \omega) \zeta_n + i \sum_m \zeta_{n-m} v_m - 4i \a v_n = 0.
\la{eq:Fourier2}
\end{align}

We expect the hidden-attractor oscillation to be dominated by the fundamental frequency $\omega$.  Taking \Eq{eq:Fourier1} for $n=0$ and \Eq{eq:Fourier2} for $n=1$, and neglecting $v_n, \zeta_n$ for all $|n| > 1$, we obtain
\begin{align}
& v_0 + \frac{1}{2i} \left( \zeta_0 - \zeta_0^\ast \right) \simeq i_{\rm tot} , \la{eq:hidden-v} \\
&[1 + i (v_0 - \omega) ] \zeta_1 \simeq i (4 \a - \zeta_0 ) v_1 . \la{eq:hidden-z}
\end{align}
The equilibrium value of $\zeta$ is, according to \Eq{eq:dimless_zeta2},
\be
\zeta_0 = \frac{4 \a}{1 - \frac{i}{v_0}} ,
\la{eq:zeta0}
\ee
where $v_0$ is the corresponding equilibrium value for $v$ (the same quantity as in \Eq{eq:deviations}).  For $|v_0| \gg 1$ we therefore have
\be
\zeta_0 \simeq 4 \a .
\la{eq:4a} 
\ee
Putting \Eq{eq:4a} into \Eq{eq:hidden-v} we get
\be
v_0 \simeq i_{\rm tot},
\la{eq:Omega0}
\ee
i.e., $v_0$ is approximately equal to the external current bias applied to the JJ.  This corresponds to the stable and approximately ohmic branch of the $I$-$V$ characteristic for the JJ.

In the regime of large applied current ($|i_{\rm tot}| \gg 1$) and small damping ($4 \a \gg 1$), Eqs.\ \eqref{eq:hidden-v} and \eqref{eq:hidden-z} are consistent with a hidden-attractor self-oscillation ($|v_1| > 0$ and $|\zeta_1| > 0$) with a fundamental frequency
\be
\omega \simeq v_0 \simeq i_{\rm tot} ~.
\la{eq:SO-f}
\ee
Note that in the original variables [see Eqs.\ \eqref{eq:charVI} and \eqref{eq:dimless_vars}], this fundamental frequency is
\begin{eqnarray}
    \Omega = \gamma \omega = \frac{2e V_0}{\hbar} = \frac{2e R I}{\hbar}.   
\end{eqnarray}
This agrees with what we expected from the nature of parametric resonance, as was explained in words in \Sec{sec:model}, after we formulated Eqs.\ \eqref{eq:MDM2-V} and \eqref{eq:MDM2-z}.  In \Sec{sec:numerical} we discuss how such hidden-attractor oscillations can be seen in numerical simulations.  There we will measure frequency in terms of an equivalent $v_0$.

\subsection{Numerical simulations}
\la{sec:numerical}

\begin{figure}[t]
	\subfigure[]{\includegraphics[width=0.45 \textwidth]{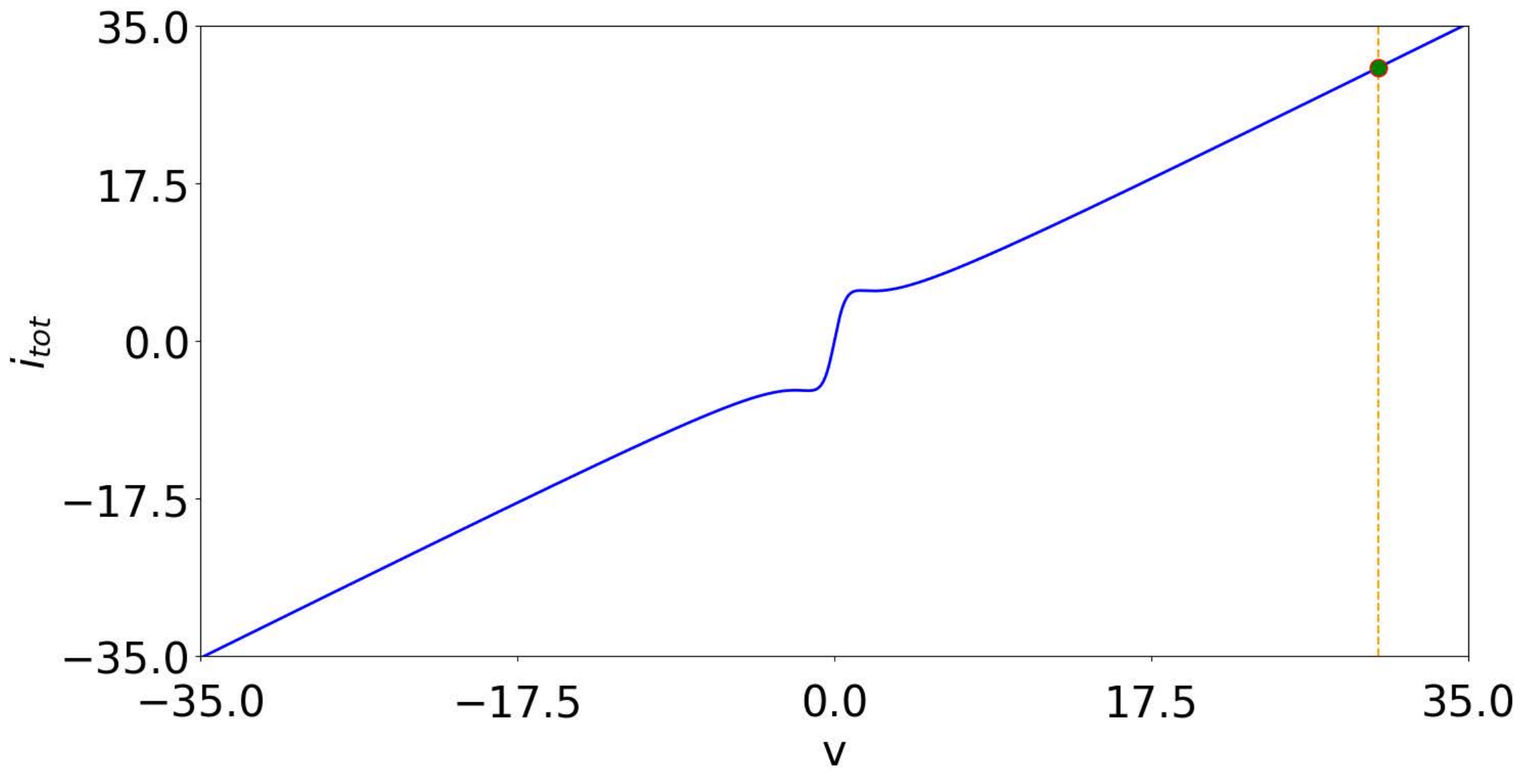}} \hskip 1 cm
	\subfigure[]{\includegraphics[width=0.45 \textwidth]{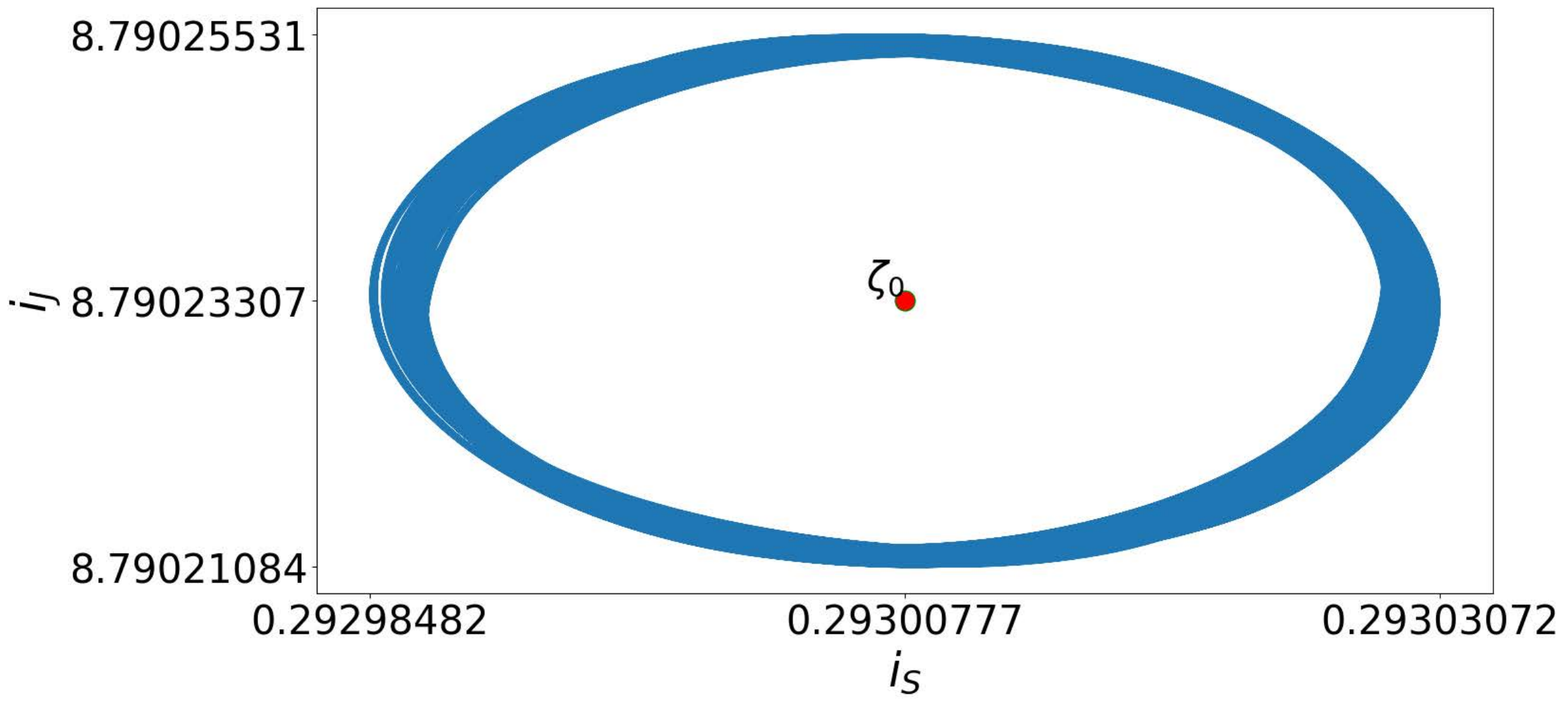}}
	\subfigure[]{\includegraphics[width=0.5 \textwidth]{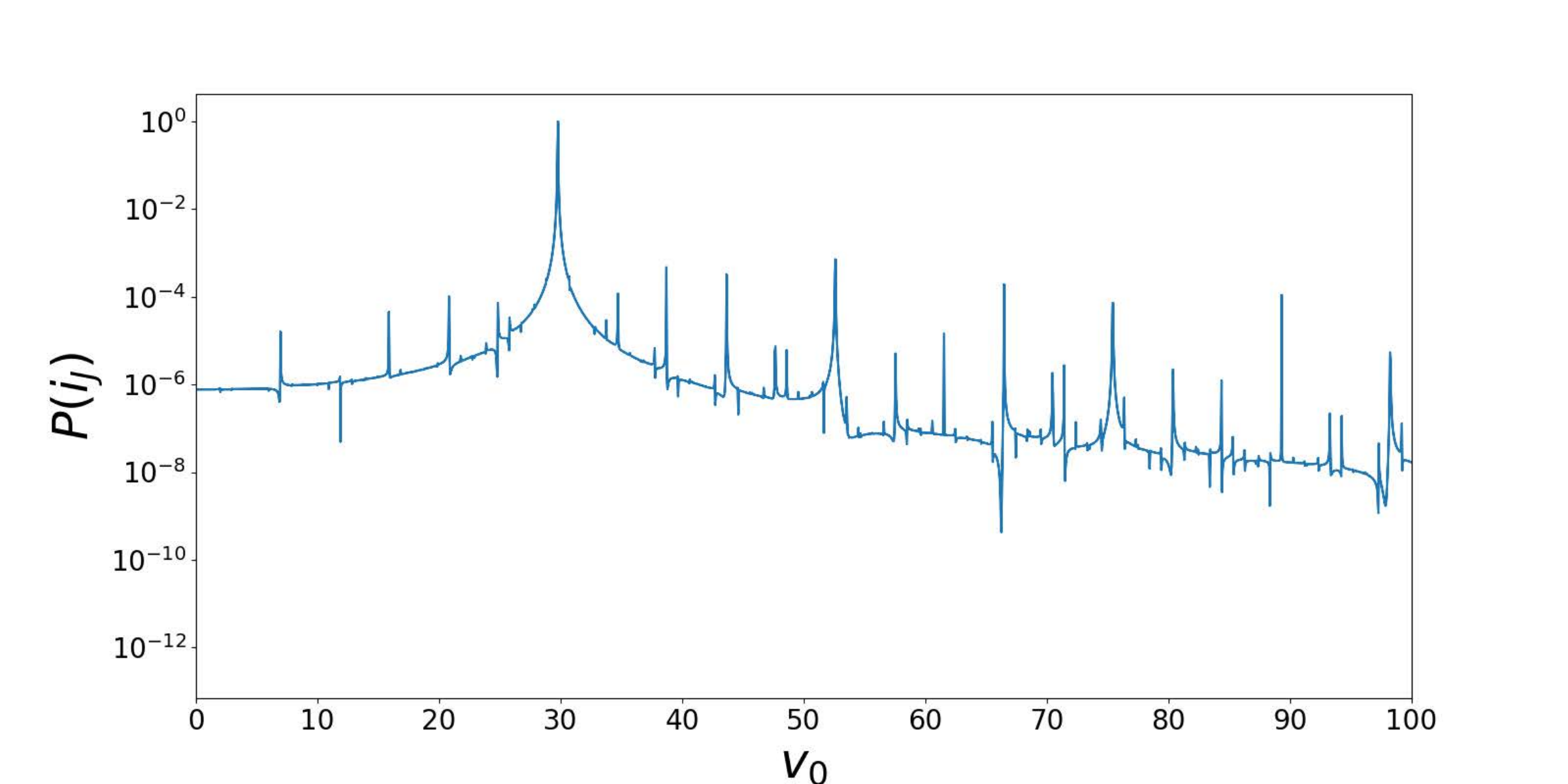}}
\caption{Hidden-attractor self-oscillation around a stable fixed point $\zeta_0$ (with $\Re \zeta_ 0 = \bar i_S$ and $\Im \zeta_0 = \bar i_J$) of the dynamical system defined by Eqs.\ \eqref{eq:dimless_zeta1} and \eqref{eq:dimless_zeta2}, for $\a = 2.2$.  (a) The initial point (in green), marked with reference to the characteristic curve (in blue) of \Eq{eq:i-v_char_dimless}.  (b) Hidden attractor in the $\zeta$ plane.  The orbit is plotted for times $500 \leq \tau \leq 550$.  The equilibrium point $\zeta_0$ is marked in orange.  (c) Power spectrum for the oscillation of the tunneling current $i_J$, computed from the numerical integration for $3500 \leq \tau \leq 3650$.\la{fig:SO}}
\end{figure}

As we already alluded to in \Sec{sec:bifurcation}, the mathematical appearance of such a self-oscillation is different from the Hopf bifurcations considered in \cite{SO, Strogatz-Hopf} and in the ``putt-putt'' engine model of \cite{engines, solar}.  The form of Eqs.\ \eqref{eq:dimless_zeta1} and \eqref{eq:dimless_zeta2} is closer to the model of the Quincke rotor in \cite{engines}, but the Quincke rotor's self-sustained rotation appeared as a simple pitchfork bifurcation, which is not the case in our model of the JJ.  As we will see, our model of the JJ is more subtle from the point of view of the mathematical theory of dynamical systems, since it exhibits hidden attractors around stable fixed points.

In order to find hidden attractors in our model and characterize some of their properties, we integrate Eqs.\ \eqref{eq:dimless_zeta1} and \eqref{eq:dimless_zeta2} numerically, for various values of the parameter $\a$ and for various choices of initial conditions.  The results shown in this section were obtained using the LSODA method, as implemented in the ODEPACK library \cite{LSODA}.  Almost identical results were obtained using the BDF method, as implemented in the CVODE library \cite{CVODE}.

As expected from the argument in \Sec{sec:hidden}, we find numerical evidence of hidden attractors around stable equilibrium points for sufficiently large values of $\a$ and $\bar v$.  To find them, we displaced the initial conditions slightly away from the equilibrium \hbox{$\delta v = \delta i_J = \delta i_S = 0$} [see \Eq{eq:deviations}] and we then observed the corresponding orbit after long times.  Figure \ref{fig:SO} shows one such hidden attractor, for an orbit with initial values $\bar v = 30$, $\delta v = \delta i_J = 0$, and $\delta i_S = -0.1$  In general, we find that hidden attractors appear only for values of $\a$ large enough that the corresponding $I$-$V$ characteristic exhibits hysteresis (see \Sec{sec:critical}).  Since the equilibrium point $\zeta_0$ inside the hidden attractor is stable, it has a finite basin of attraction.  For the same parameters used in the simulation of \Fig{fig:SO} but with initial $|\delta i_S| \lesssim 10^{-6}$ (while keeping the initial $\delta v = \delta i_J = 0$), we find that the resulting orbit decays to $\zeta_0$, rather than giving a persistent oscillation.

Thus, a small but finite initial disturbance is needed to excite the JJ's self-oscillation.  This is the defining feature of a ``hidden attractor''.  The power spectrum for $i_J$ shows that the hidden attractor corresponds to an oscillation whose fundamental frequency is $v_0$ (in the dimensionless variables defined in \Eq{eq:dimless_vars}).  This relation between the fundamental frequency of the hidden-attractor oscillation and the applied voltage $v_0$ holds over a range of values of $v_0$, as shown in \Fig{fig:v-freq}.  This result is consistent with the fact that the electromagnetic and phonon radiation from a DC-biased JJ has fundamental frequency $\Omega = 2 e V / \hbar$.  Note, however, that in our model the simple formula of \Eq{eq:AC} does not hold.  The conversion between the applied DC bias and the observed AC emission (what we called \hbox{``DC $\to$ AC''} in \Sec{sec:intro}) appears here as the result of an irreversible engine dynamics, associated with the presence of hidden attractors in the solutions to our equations of motion.

\begin{figure}[t]
	\includegraphics[width=0.45 \textwidth]{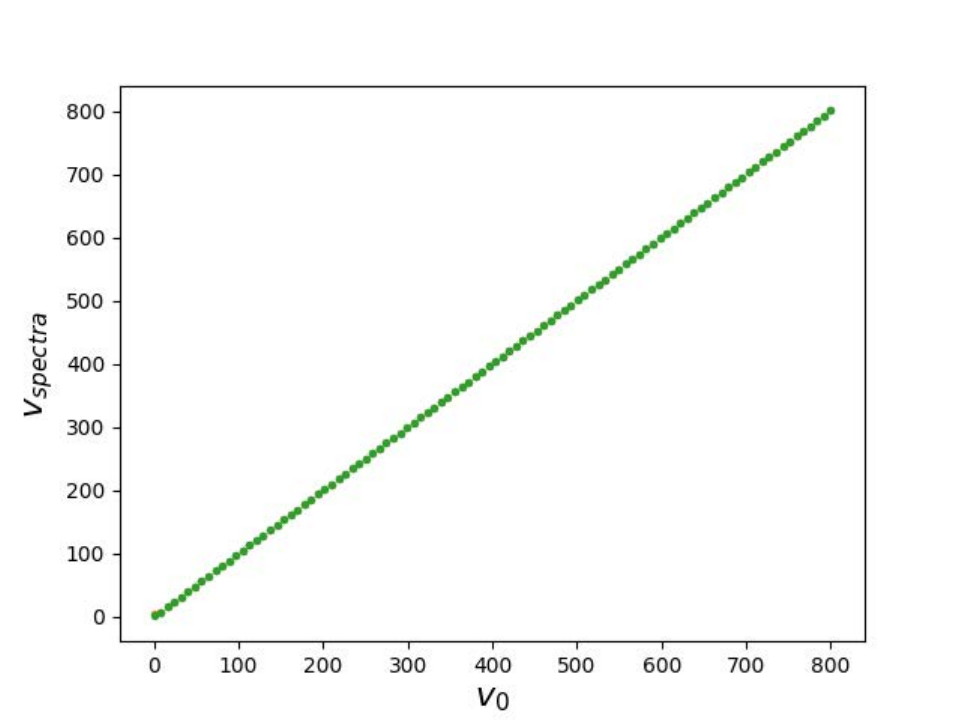}
\caption{Value of the fundamental peak of the frequency spectrum for the self-oscillation of the tunneling current $i_J$ [labelled $v_{\rm spectra}$ and measured in the units defined in Eqs.\ \eqref{eq:charVI}, \eqref{eq:dimless_vars}] as a function of the applied voltage $v_0$, for $\a=2.2$.  The spectra are computed from the numerically integrated orbits for $4850 \leq \tau \leq 5000$.\la{fig:v-freq}}
\end{figure}

For $\a < 2$ ---in which case the $I$-$V$ characteristic for the JJ shows no hysteresis--- we have not found hidden attractors in our numerical simulations.  Instead, we find that initial conditions close to the linearly stable equilibrium decay steadily towards that equilibrium.  An example of such behavior is shown in \Fig{fig:noSO}.  A more precise and extensive characterization of the dynamical system defined by \Eq{eq:dimless_set} is left for future investigation, as it constitutes and interesting mathematical problem in its own right.  Here our goals has been only to establish the presence of hidden attractors whose behavior is consistent with the \hbox{DC $\to$ AC} properties of the Josephson effect.

We consider it non-trivial that our simple model of the JJ gives an adequate picture not only of the unusual properties of the $I$-$V$ characteristic (as explained in \Sec{sec:stab}) and of its response to external driving by an AC electric field (as explained in \Sec{sec:Shapiro}), but also of its autonomous operation as an engine (self-oscillator) that acts as an irreversible \hbox{DC $\to$ AC} converter.  We believe that this calls for further investigation not only in terms of the physical theory of open quantum systems, but also of the mathematical theory of dynamical systems.  The aim should be to clarify conceptually the relation between these different phenomena, all of which are characteristic of an active electronic device.

The numerical study of hidden attractors is a rather subtle problem that deserves a deeper investigation in the mathematical theory of nonlinear dynamical systems.  For our present purposes it suffices to show that self-oscillations persist over very long times: we observed this for $\tau$ as large as $5000$.  We were able to establish this using both the CVODE and the ODEPACK libraries to carry out our numerical integration.

Since the ODEs for our model might be stiff (see, e.g., \cite{numerics}), it is important to establish that the self-oscillations observed are not due to instabilities of the integration method.  We therefore solved the equations using various different stable methods, including the CVODE and ODEPACK libraries already mentioned, and compared the results.  All of the methods that we used gave long-lived self-oscillations with the correct Josephson frequency.  However, these self-oscillations did not appear to be infinitely long lived (i.e., hidden attractors) in simulations with some other commonly used numerical methods.  In particular, as shown in \Fig{fig:numerics}, for the same initial conditions as in \Fig{fig:SO}, the SUNDIALS library \cite{SUNDIALS} (which uses the Radau method) approaches the stable fixed point.

Comparing the LSODA and Radau trajectories shown in \Fig{fig:numerics}, we find that the former (which seems to show a hidden attractor) is much smoother than the latter (which does not).  We are therefore confident that the former is more likely to reflect the correct long-term behavior of our dynamical system of interest.  The trouble with the Radau solution of \Fig{fig:numerics} seems to be related to accumulation of round-off error and to the time steps being too large.  This method's implementation varies adaptively between 5th and 13th order.  The sensibility at different time steps may be misestimated, leading to a wrong interpolation.

\begin{figure}[t]
	\subfigure[]{\includegraphics[width=0.45 \textwidth]{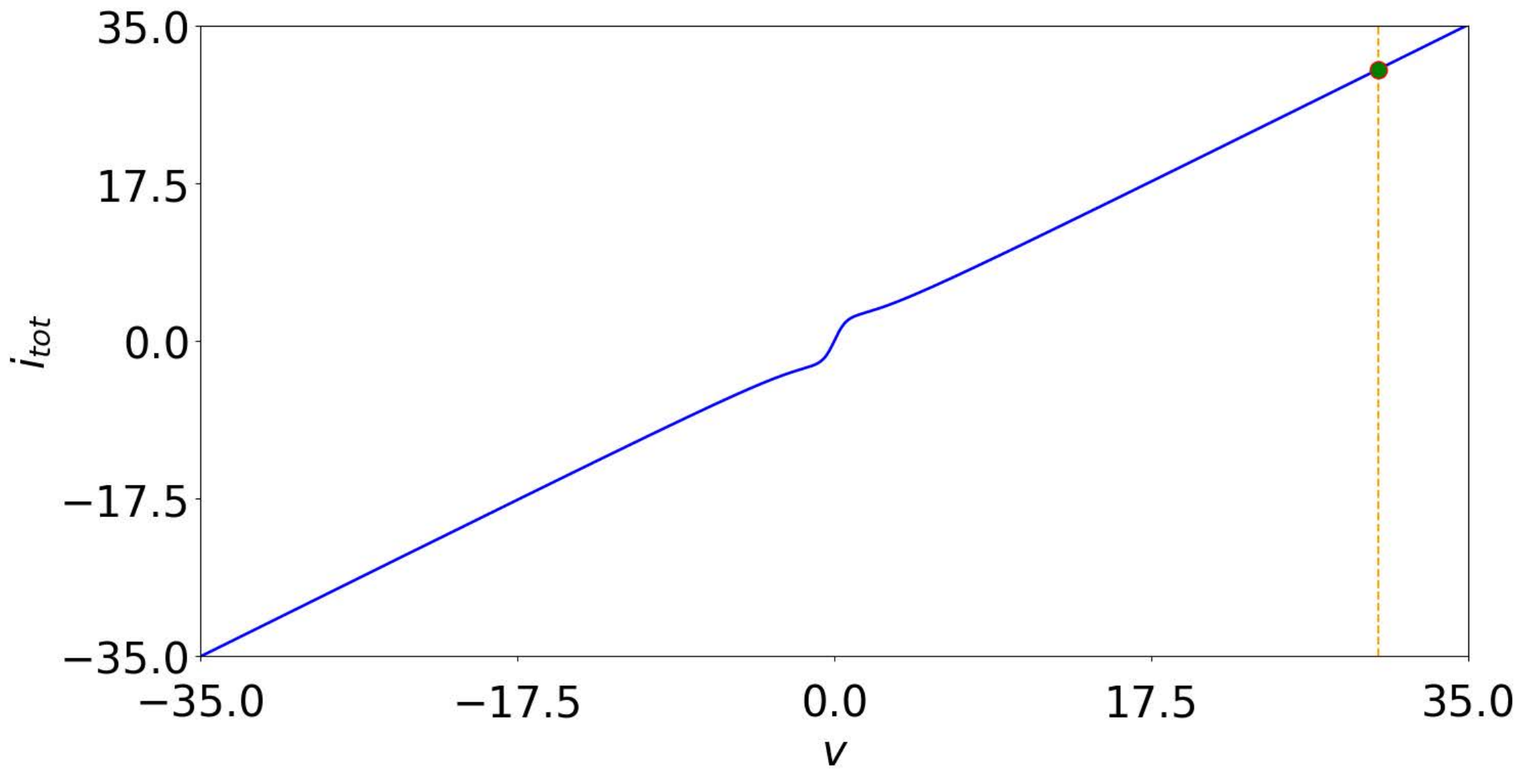}} \hskip 1 cm
	\subfigure[]{\includegraphics[width=0.45 \textwidth]{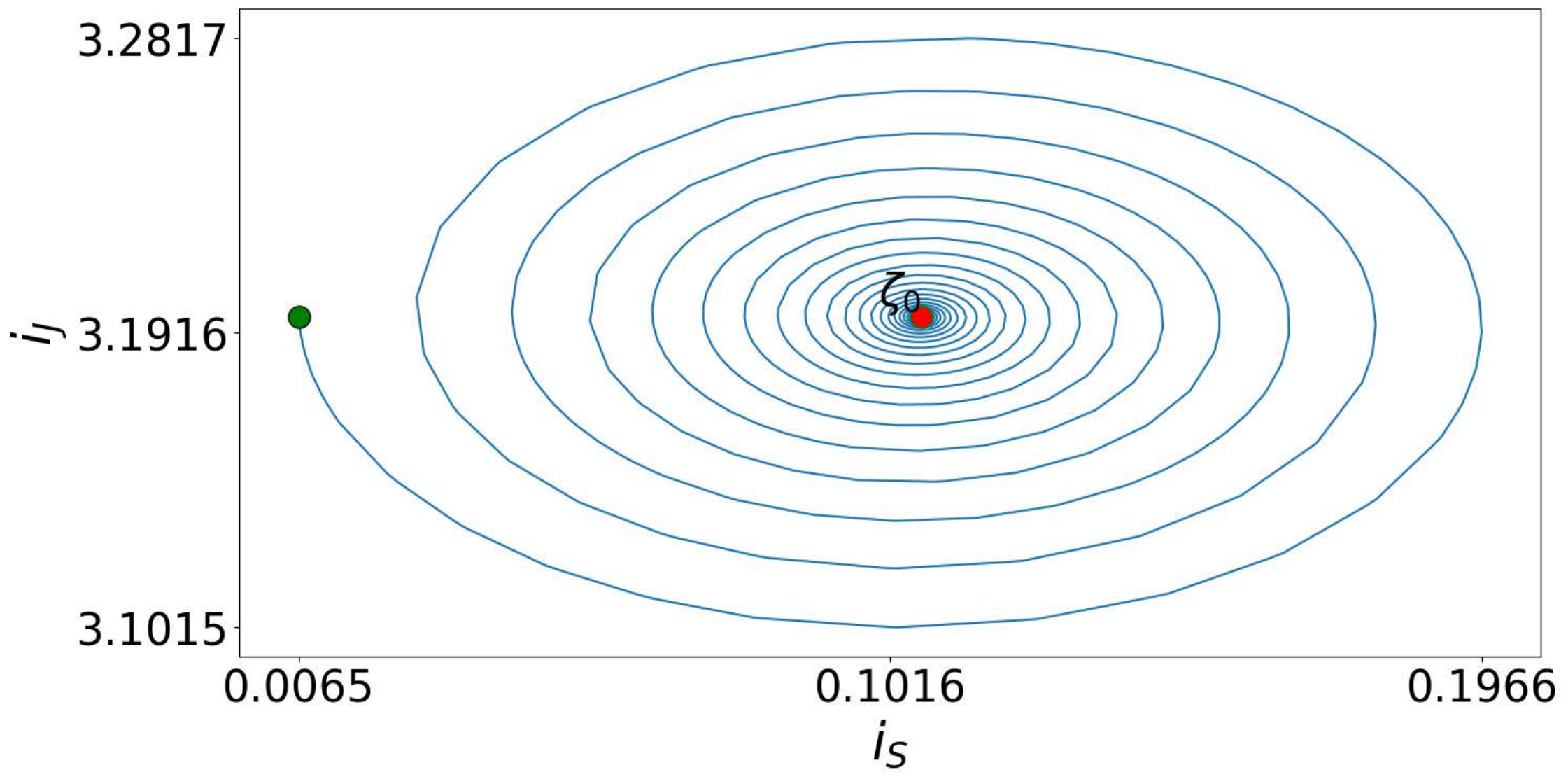}}
\caption{Approach to a stable fixed point of the dynamical system defined by \Eq{eq:dimless_set} for $\a = 0.8$.  (a) The initial point, marked in green with reference to the characteristic curve (in blue) of \Eq{eq:i-v_char_dimless}. (b) Orbit in the complex $\zeta$ plane for $0 \leq \tau \leq 550$.\la{fig:noSO}}
\end{figure}

Another interesting problem raised by our model concerns the possibility of chaos (strange attractors), both in the homogeneous system of \Eq{eq:dimless_set} and in the inhomogeneous system with an external periodic driving of \Eq{eq:dimless_set_forced}.  Although we must leave this for future investigation, here we make some general remarks about this problem.

In van der Pol's simple model and related formulations of self-oscillation (see \cite{SO} and references therein), the total phase-space for the homogeneous system has only two dimensions and (by the Poincar\'e-Bendixson theorem) strange attractors are not possible; see, e.g., \cite{Strogatz-PB}.  This is also true of the RCSJ model as described by \Eq{eq:WBeom}.  However, as we stressed in \Sec{sec:feedback}, these models are not physically realistic because they invoke a negative linear damping, rather than describing the active dynamics of the self-oscillators in terms a feedback between the oscillator and another degree of of freedom.  If that degree of freedom is added, the Poincar\'e-Bendixson theorem becomes inapplicable.  Thus, e.g., the presence of both limit cycles and strange attractors has been shown numerically for the homogenous ``leaking elastic capacitor'' model, which describes a classical engine based on feedback between a mechanical and an electrical degree of freedom \cite{LEC}.

On the other hand, one of the earliest results in what would later be called ``chaos theory'' was the work of Cartwright and Littlewood on the forced (inhomogenous) van der Pol equation (see \cite{Guckenheimer, Bolt-etal} and references therein).  The appearance of chaos is associated with the presence of a {\it quasiperiodic} regime in which the ratio of the frequency of the external driving and the natural frequency of the self-oscillator is irrational.  An interesting open question is whether something similar occurs for the inhomogeneous \Eq{eq:dimless_set_forced} and whether this might have something to do with the structure of the Shapiro steps that we obtained in \Sec{sec:Shapiro}.

So far we have not identified any chaotic regimes for Eqs.\ \eqref{eq:dimless_set} or \eqref{eq:dimless_set_forced}.  We expect that the investigation of that problem will require a better handle on the numerical methods suited to the detailed study of hidden attractors in general and of our dynamical model of the Josephson effect in particular.  Identifying attractors in numerical simulations of open systems is a non-trivial problem that may be of considerable physical interest (see, e.g. \cite{optomechanical}).

\begin{figure}[t]
	\includegraphics[width=0.45 \textwidth]{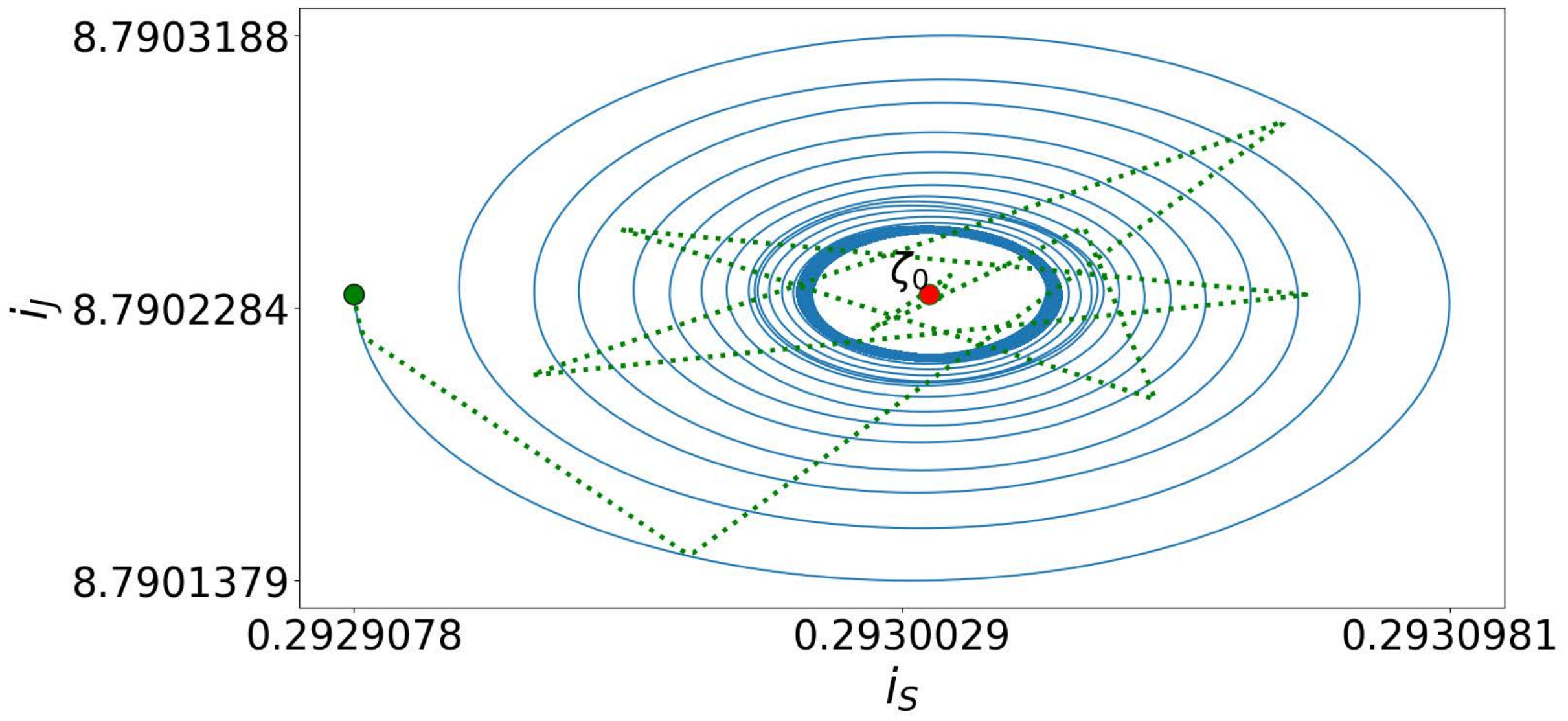}
\caption{Integrated orbit found using the LSODA (in blue) and Radau (dotted green).  The green dot denotes the initial point, while the orange one denotes the fixed point $\zeta_0$. This curve has the same parameters as the \Fig{fig:SO}, but with initial condition $\delta i_S = - 1 \times 10^{-4}$.  The hidden attractor is only observed in the LSODA integration\la{fig:numerics}} 
\end{figure}

\section{Electromagnetic radiation}
\la{sec:radiation}

Self-oscillation of the JJ produces both electromagnetic and acoustic waves, with fundamental frequency $\Omega = eV/\hbar$ and weaker harmonics. In the electromagnetic case, the corresponding wavelength is much longer than the JJ's linear dimension, suggesting a collective character of the emission process. Moreover, since Cooper pairs are (approximately) bosons, their wave function is symmetric under permutations.  This makes our model of the JJ well suited for implementing Dicke's superradiance mechanism \cite{Dicke}. Such radiation is highly monochromatic and coherent, and when the JJ is placed inside a resonant cavity the emission is considerably enhanced, as shown experimentally in \cite{Pedersen, laser}.

\subsection{Superradiance dynamics}
\la{sec:superMDM}

Here we will briefly discuss the phenomenon of superradiance for an idealized system of $N$ two-level ``atoms'' interacting with an electromagnetic field at zero temperature.  We assume that the size of the sample is small compared to the wavelength of emitted radiation, leading to collective emission of radiation with intensity proportional to $N^2$. If the initial state of the atomic ensemble is a product of $N$ identical $2 \times 2$ density matrices, one can show that in the limit of large $N$ the product structure is preserved and that the time evolution of the system may be described by the non-linear master equation for the corresponding MDM, normalized to $N$:
\be
\frac{d}{dt}\sigma = - i \omega_A [S_3 , \sigma] + \frac 1 2 \gamma_e \left[ {\rm Tr} (\sigma S^+) S^- - {\rm Tr} (\sigma S^-) S^+ , \sigma\right] .
\label{eq:nonlME}
\ee
For details on this formulation, see \cite{AL}.

Here, $\omega_A$ is an atomic frequency and $\gamma_e$ is the spontaneous emission rate for the single atom.  Denoting by $|1\rangle , |2\rangle$ the excited and ground state respectively, we define the ``spin'' matrices by
\be
S_3 |1\rangle = \frac 1 2 |1 \rangle, \quad S_3|2 \rangle = -\frac 1 2 |2 \rangle, \quad S^+ |2 \rangle = |1 \rangle, \quad \hbox{and} \quad S^- |1\rangle = |2 \rangle .
\la{eq:spins}
\ee 
Inserting, the same parametrization of the MDM in terms of $n_{1,2}$ and $z$ as in \Eq{eq:MDMapp} into \Eq{eq:nonlME}, we obtain
\bea
\dot n_1 &=& - \gamma_e |z|^2 , \quad \dot n_2 = \gamma_e |z|^2 , \nl
\dot z &=& - i \omega_A  z +  \frac 1 2 \gamma_e(n_1-  n_2) z .
\la{eq:MDMeqAp1}
\eea

Because the total number of atoms $N = n_1 + n_2$ is preserved, we can rewrite this in terms of the variable $n = n_1 - n_2$ as
\bea
\dot n &=& - 2\gamma_e  |z|^2 ,   \nl
\dot z &=& - i \omega_A z -  \frac 1 2 \gamma_e n z .
\label{eq:MDMeqAp}
\eea
Note that, due to the mathematical structure of \Eq{eq:nonlME}, the initially pure MDM remains pure (i.e., $|z|^2 = n_1 (N-n_1)$) and the evolution of excited state population satisfies the well known equation
\be
\dot n_1=- \gamma_e (N- n_1) n_1 .
\la{superradiance}
\ee
This reproduces the $N^2$ proportionality and the bell-shaped time dependence of the intensity of radiated energy seen in superradiance.  The radiated power is given by
\be
{\cal P}(t) = -\hbar \omega_A \dot n_1 = \hbar \gamma_e \omega_A |z(t)|^2 .
\label{eq:power}
\ee

Note that it is not easy to implement this superradiance effect with actual atoms (see, e.g., \cite{Gross-Haroche}).  This requires that the distance between the atoms be much smaller than the radiation's wavelength and also that the atoms occupy a symmetric configuration in space.  Only under those two conditions would it be possible to express the interaction of the atoms with the electromagnetic field as the sum of dipoles multiplied by a single field operator.  However, in the case of the JJ the two levels correspond to the left-hand and right-hand electrodes, while the transition between levels corresponds to tunneling of a Cooper pair across the junction. In this case, the two conditions for superradiance listed above are satisfied automatically.  First, the linear size of the junction is of the order of micrometers, while the observed radiation is typically submillimeter.  Second, Cooper pairs are (to a very good approximation) non-interacting bosons whose state is symmetric with respect to exchanging the two electrodes.  Although the superradiance model was introduced in a very different context, it is therefore ideally suited for describing the production of non-thermal radiation by the JJ.

To calculate the efficiency with which the JJ transforms the applied external power into radiation, we may add to the JJ equations of motion for $V$ and $z$ [Eqs.\ \eqref{eq:MDM2-V} and \eqref{eq:MDM2-z}] the superradiant contributions resulting of \Eq{eq:MDMeqAp}, recalling that $V = e n/C $ [\Eq{eq:V}]. The modified JJ dynamics is then governed by
\bea
\dot V &=&- \gamma V  + i \frac{2Ke}{C} (z - z^\ast) + \frac 1 C \left(I - 2e\gamma_e  |z|^2 \right), \la{eq:MDMApp-V} \\
\dot z &=& -\left( \gamma + \gamma_e \frac{CV}{2e} \right) z - i \frac{2 e V}{\hbar} z + i K \frac{CV}{e} . \la{eq:MDMApp-z}
\eea
The correction $-2e\gamma_e  |z|^2$ to the external current $I$ in \Eq{eq:MDMApp-V} corresponds, when multiplied by the voltage $V$, to the radiated power.  We can therefore introduce a parameter characterizing the JJ's radiation efficiency under stationary conditions:
\be
\eta_{\rm rad} =  \frac{\mathcal{P}_{\rm rad}}{\mathcal{P}_{\rm in}}= \frac{ 2e V \gamma_e  |z_0|^2}{VI} = \frac{2e \gamma_e  |z_0|^2}{I} ,
\la{eq:eff}
\ee
where $z_0$ is the stationary value of $z$ in Eqs.\ \eqref{eq:MDMApp-V} and \eqref{eq:MDMApp-z}.  Note that the radiated power $2eV \gamma_e  |z_0|^2$ computed from \Eq{eq:MDMApp-V} agrees with the expression of \Eq{eq:power} if we take an ``atomic frequency'' $\omega_A = 2 eV / \hbar$.

Rather than using the dimensionless variables that we introduced in \Sec{sec:open} [see \Eq{eq:dimless_vars}], here it will be more convenient to work with the angular frequency
\be
\Omega = \frac{2eV}{\hbar}
\ee
and the Cooper-pair number current
\be
J \equiv \frac{I}{2e} .
\ee
Introducing the frequency parameter
\be
\omega_c \equiv \frac{4e^2}{\hbar C} ,
\la{eq:omegac}
\ee
Eqs.\ \eqref{eq:MDMApp-V} and \eqref{eq:MDMApp-z} take the form
\bea
\dot \Omega &=& - \gamma \Omega  + i {K \omega_c} (z - z^\ast) + \omega_c \left( J - \gamma_e |z|^2 \right) \nonumber \\
\dot z &=& - \left( i \Omega + \gamma + \gamma_e \frac{\Omega}{\omega_c} \right) z +i\frac{2K}{\omega_c} \Omega.
\la{eq:MDM6}
\eea
The stationary solution $z = z_0 = $ const., $\Omega = \Omega_0 =$ const.\ corresponds to
\be
z_0 = \frac{2K}{\omega_c} \left[ \frac{\Omega_0^2 + i (\gamma+ \gamma_e \Omega_0 / \omega_c ) \Omega_0}{\Omega_0^2 + ( \gamma + \gamma_e \Omega_0 / \omega_c )^2} \right] .
\la{eq:z0}
\ee
In the weakly damped regime (which, by the results of \Sec{sec:SO}, is that one in which we expect a self-oscillating dipole) we have that 
\be
|z_0|^2 \simeq \frac{4 K^2}{\omega_c^2} \simeq \frac{I_c}{e \omega_c} ,
\la{eq:z02}
\ee
where $I_c$ is the physical value of the critical Josephson current [see  \Eq{eq:Ic0}].  Therefore, combining Eqs.\ \eqref{eq:eff} and \eqref{eq:z02}, we find that the radiation efficiency of the weakly damped JJ is:
\be
\eta_{\rm rad} = \frac{2 \gamma_e}{\omega_c} \frac{I_c}{I} = \gamma_e \frac{\hbar C}{2 e^2} \frac{I_c}{I} .
\la{eq:eff-I}
\ee

\subsection{Radiation into open space}
\la{sec:space}

For a single JJ that radiates microwaves into open space, we can use the expression for $\gamma_e$ computed for a two-level atom with an electric dipole $\bf d$.  This is given by the formula
\be
\gamma_e = \frac{4\pi\mu_0}{3\hbar c}\mathbf{d}^2 \omega_A^3 .
\la{eq:gammae}
\ee
where now
\be
\omega_A = \frac{2eV}{\hbar} ,
\la{frequency}
\ee
and where the dipole moment of a ``JJ  atom'' is given by the product of the distance $\ell$ between the superconducting electrodes and the charge $2e$ for a single Cooper pair. This leads to the final formula for $\eta_{\rm rad}$, valid in the weak damping regime:
\be
\eta_{\rm rad} = \frac{I_c}{I} \frac{64 \pi \mu_0 e^3}{3 \hbar^3 c} C V^3 \ell^2 .
\la{efficiency2}
\ee
Typically, radiation is observed from JJ's in which $I \simeq I_c \simeq 10^{-3}$ A,  $C \simeq 3 \times 10^{-13}$ F,  $V \simeq 10^{-3}$ V, and $\ell =  10^{-9}$ m, which gives
\be
\eta_{\rm rad} \simeq 10^{-7} .
\la{eq:eff-n}
\ee
This efficiency can be enhanced by placing the JJ inside a high-quality optical cavity, as we will discuss in \Sec{sec:cavity}.  The resulting estimate of the efficiency is reasonable, but a question remains because the coherence length $\xi_0$ of the Cooper pairs may be much larger than the separation $\ell$ between the electrodes \cite{Delin}.  The proper interpretation of the dipole moment of the ``JJ atom'' could therefore require more careful consideration, an issue that we must leave for future research.

\subsection{Cavity enhancement}
\la{sec:cavity}

In the experimental setup of \cite{Pedersen, laser}, the JJ is placed inside a resonant cavity.  This leads to much stronger and highly coherent radiation, which the authors of \cite{laser} call a ``Josephson junction laser''.  To estimate the efficiency in this case, we use a well-known result of cavity quantum electrodynamics (QED) that describes enhanced spontaneous emission by an atom in resonance (the ``Purcell effect'' \cite{Purcell}):
\be
\gamma_e^{\rm cav} = \gamma_e \frac{3Q}{4\pi^2} \left( \frac{\lambda_A}{L}\right)^3 ,
\la{eq:gamma_cav}
\ee
where $Q$ is a quality factor of the cavity, $L$ is a cavity linear dimension, and $\lambda_A = 2\pi c/ \omega_A$ is the wavelength of the emitted radiation.  Typically $L \simeq\lambda_A$, so that the enhancement of the efficiency factor is largely determined by $Q$.  Therefore we expect that
\be
\eta_{\rm rad}^{\rm cav} \simeq 0.1 \, \eta_{\rm rad} Q ,
\la{eq:eff-Q}
\ee
which is consistent with the observed enhancement by a few orders of magnitude, relative to \Eq{eq:eff-n}.  For further treatment of the Purcell effect in cavity QED, see \cite{cavities} and references therein.

\section{Discussion}
\la{sec:discussion}

Despite having been amply studied and applied since it was first proposed in 1962, the Josephson effect has remained something of a conundrum because of the failure of theorists to treat it consistently as the dynamics of an open system.  The JJ's ability to convert DC$\to$AC should not be regarded as a spontaneous breaking of time-translation invariance (in the manner of a ``time crystal'' \cite{crystal}), but rather as the dynamics of an engine that cyclically extracts work from an external disequilibrium.  This DC$\to$AC conversion is a {\it self-oscillation}, and as such a thermodynamically irreversible process resulting from a feedback between the state of a ``tool'' (corresponding, in this case, to the coherence $z$ in the macroscopic density matrix of \Eq{eq:sigma}) and the coupling of the engine's working substance (here the Cooper pairs in the JJ) to the external baths.  In our model, this feedback is captured by the coupling of the differential equations for the evolution of $z$ and the voltage $V$ of the junction [Eqs.\ \eqref{eq:MDM2-V} and \eqref{eq:MDM2-z}].

We have shown that the dynamical system characterized by the coupled Eqs.\ \eqref{eq:MDM2-V} and \eqref{eq:MDM2-z} has a rich structure, consistent with the key features of the JJ.  One of these features is the voltage-current relation of \Eq{eq:IVrelation}, which exhibits ohmic behavior for large voltage bias, as well as negative differential resistance and hysteresis when the damping of the system is small enough.  In \Sec{sec:characteristic} we recovered the main qualitative results of the RCSJ model, but without invoking a ``tilted washboard potential'' unbounded from below, or introducing any {\it ad hoc} dissipation mechanism extraneous to the JJ's dynamics.  Note also that the current-voltage characteristic for the JJ, as shown in \Fig{fig:IV}(a) displays a ``pinched'' hysteresis loop (i.e., a closed hysteresis curve that passes through the origin, at zero current and zero voltage), meeting Chua's updated definition of a ``memristor'' \cite{Chua}.  Our model might therefore be useful in clarifying the debate about whether memristors are passive or active devices \cite{memristor1, memristor2}.

Our model of the JJ can describe the well known quantum interference used in SQUIDs for the precise measurement of magnetic fluxes, as we showed in \Sec{sec:squid}.  Another important feature of our treatment of the JJ as an open system is that it accounts for the frequency locking (``Shapiro steps'') observed when the JJ is subjected to an external AC voltage driving, as follows directly from the electromagnetic irradiation used in most experimental setups.  As we discussed in \Sec{sec:Shapiro}, this effect emerges naturally in our description, without having to invoke any accompanying modulation of the external current (as must be done in the RCSJ model).

In \Sec{sec:SO} we found by numerical integration that this same dynamical system exhibits self-oscillations for sufficiently large external DC bias, in the form of ``hidden attractors'' that can be excited by small but finite perturbations about a linearly stable equilibrium.  This self-oscillation of $z$ and $V$ has a fundamental frequency equal to the Josephson frequency $\Omega = 2 e V/ \hbar$, and can therefore account for the emission of monochromatic photon and phonon radiation observed in many experiments (a radiation which, in the case of photons, may be enhanced by the coupling of the JJ to a resonant cavity).  This monochromatic radiation can be understood as a form of work extracted by the JJ engine from the disequilibrium between the two external electronic baths to which it is coupled.  This is consistent with the picture of JJ radiation presented in \Sec{sec:radiation}, based on Dicke's theory of superradiance in quantum optics.  However, the precise relation between the hidden attractors of \Sec{sec:SO} and the superradiant mechanism discussed in \Sec{sec:radiation} calls for further investigation.

The model that we have presented here is the simplest dynamical description consistent with these key features of the JJ.  Further work is needed to improve and clarify the detailed modelling of the JJ, including such effects as the dependence of the tunneling rate $K$ on the junction voltage $V$ and an explicit treatment of the electrostatic interactions of the Cooper pairs.  The methods that we have introduced here might also help to clarify the dynamics of the Esaki diode, an active device used to build electrical self-oscillators that, like the JJ, exhibits a negative differential resistance associated with electron tunneling \cite{Esaki}.

Beyond the concrete problem of understanding the Josephson effect as an irreversible process, we consider that the present work is also of interest more broadly to the physical theory of quantum thermodynamics as well as to the mathematical theory of dynamical systems.  In previous models of quantum engines, the work is extracted by a degree of freedom that is effectively classical or semi-classical \cite{QT}.  In our model of the JJ work is extracted by a ``tool'' (or ``piston'') that is a macroscopic but intrinsically quantum object: the coherence $z$ in the density matrix of \Eq{eq:sigma}.  This could make it important to developing a better understanding the physics of practical qubits and how the can be maintained against decoherence.

In the theory of dynamical systems, most mathematical models of autonomous engines have been based on Hopf bifurcations and the associated limit cycles.  In our model, the JJ's engine dynamics is associated with hidden attractors \cite{hidden}.  The precise characterization of these attractors and their relation to other features of our model deserves further study.  Following the suggestion made long ago by Fabry and Le Corbeiller \cite{LeCorbeiller-Eng, 2Stroke}, we believe that a fruitful dialogue between thermodynamics and the theory of differential equations may be the key to a better understanding of active devices \cite{passive-active}, including the JJ and other electrical self-oscillators.

\begin{acknowledgments} We thank Luis Cort-Barrada for extensive discussions and valuable suggestions. This work was supported by the International Research Agendas Programme (IRAP) of the Foundation for Polish Science (FNP), with structural funds from the European Union (EU). MH and GS acknowledge the support by the Polish National Science Centre's Grant OPUS--21 (No.\ 2021/41/B/ST2/03207). MH was also partially supported by the QuantERA II Programme (No.\ 2021/03/Y/ST2/00178, acronym ExTRaQT), with funds from the EU's Horizon 2020 programme. \end{acknowledgments}



\end{document}